\documentclass[AMA,Times1COL]{WileyNJDv5}
\usepackage{amsthm}
\theoremstyle{remark}
\newtheorem*{justification}{Justification}
\usepackage{tikz}
\usetikzlibrary{shapes, arrows, positioning}
\definecolor{darkgreen}{HTML}{006400}

\articletype{}%
\usepackage{subcaption}
\usepackage{colortbl} 
\received{Date Month Year}
\revised{Date Month Year}
\accepted{Date Month Year}
\journal{}
\volume{00}
\copyyear{2023}
\startpage{1}

\makeatletter
\gdef\@history@dates{}%
\gdef\@DOI@text{}%
\gdef\@doi{}%
\def\oddfoot@titlepage@info{%
  \vbox{\hsize\textwidth{%
    \noindent\rule{\textwidth}{.5pt}\vskip-40pt
    {\fontsize{7}{8}\selectfont\hfill{\pagenumfont\@FirstPg}}%
    \vskip10pt}}}%
\def\evenfoot@titlepage@info{\oddfoot@titlepage@info}%
\gdef\@jnlcitation{}%
\makeatother

\begin{document}

\title{Decoupling risk and masking in mammographic density under irregular follow-up using a latent Markov progression–detection framework}

\author[1,2]{Furkan Danisman}

\author[3]{Zarina Oflaz}

\author[4]{Zeynep Kalaylioglu}

\author[5]{Mahmut Onur Kulturoglu}

\author[5]{Lutfi Dogan}




\titlemark{}

\address[1]{\orgdiv{Department of Statistical Sciences}, \orgname{University of Toronto}, \orgaddress{\city{Toronto}, \state{Ontario}, \country{Canada}}}

\address[2]{\orgname{Vector Institute for Artificial Intelligence}, \orgaddress{\city{Toronto}, \state{Ontario}, \country{Canada}}}

\address[3]{\orgdiv{Department of Mathematics and Statistics}, \orgname{Qatar University}, \orgaddress{\city{Doha}, \country{Qatar}}}

\address[4]{\orgdiv{Department of Statistics}, \orgname{Middle East Technical University}, \orgaddress{\city{Ankara}, \country{Türkiye}}}

\address[5]{\orgdiv{Department of Surgical Oncology}, \orgname{Etlik City Hospital Oncology}, \orgaddress{\city{Ankara}, \country{Türkiye}}}

\corres{Corresponding author Furkan Danisman. \email{furkan.danisman@mail.utoronto.ca}}


\abstract[Abstract]{Background: Mammographic density is a strong marker of breast cancer risk, yet it also reduces mammographic sensitivity through masking. Screening densities are observed at irregular times, requiring methods that accommodate irregular follow-up.

Methods: We propose a two-phase framework for irregular mammographic screening data. After regularizing the irregular density histories, the first phase fits a latent Markov model within each BMI group, under the assumption that density does not itself drive risk but the underlying latent disease process does. Extracting the latent-state information leaves density to act only as a detectability factor, and the second phase links the latent process and the final density to cancer diagnosis through a detection--risk model. Detection probabilities are treated as sensitivity parameters to quantify how the estimated risk changes under different masking assumptions. Uncertainty in the inferred latent states is quantified via posterior path sampling and bootstrapping.

Results: In 616 patients, higher parity was associated with faster movement toward lower-risk, while a family history of breast cancer was associated with longer persistence in higher-risk. Moving from ignoring masking to accounting for masking with empirically supported detectability scenarios raised the estimated breast cancer risk by 70\% for post-menopausal overweight patients and about 40\% for obese patients regardless of menopausal status.

Conclusions: By attributing risk to latent progression while letting observed density operate through
detectability, the framework decouples risk from detectability, quantifies how much masking can distort breast cancer risk, and provides interpretable risk characterization under irregular
screening follow-up.}

    
\jnlcitation{\cname{%
\author{Danisman, F.}, 
\author{Oflaz, Z.}, 
\author{Kalaylioglu, Z.}, and
\author{Lutfi Dogan}} 
\ctitle{Decoupling risk and masking in mammographic density under irregular follow-up using a latent Markov progression–detection framework} 
\cjournal{\it Statistics in Medicine.} \cvol{2025;00(00):1--18}.}

\maketitle

\section{Introduction}\label{sec:introduction}

Breast cancer remains a major public health issue and a leading cause of cancer morbidity and mortality among women \cite{0.2,0.3,0.4}. Recent global estimates indicate that female breast cancer has surpassed lung cancer as the most commonly diagnosed cancer worldwide, with approximately 2.3 million new cases, representing 11.7\% of all new cancer diagnoses compared with 11.4\% for lung cancer \cite{0.3}. Over the same period, cancer accounted for nearly 10 million deaths globally \cite{0.4}. Hence, improving risk assessment and screening strategies to support earlier detection and timely intervention is crucial \cite{0.2}. 

Mammographic breast density is one of the strongest imaging-derived indicators associated with breast cancer risk and is routinely available in screening practice \cite{3.6}. A substantial literature has established density as an independent risk marker \cite{3.1,3.2,3.3,3.7,3.8,3.10} and has further shown that changes in density over time carry additional information about risk \cite{3.4}. Density assessments are comparatively inexpensive and widely collected, making them a significant longitudinal signal for risk evaluation \cite{4.2}. However, density has a dual role that complicates interpretation. Dense tissue is associated with elevated underlying risk while also inducing a masking effect that reduces mammographic sensitivity, particularly in very dense breasts \cite{4.1,4.3,4.4,4.5,4.6}. Several approaches have proposed supplemental imaging for women with dense breasts to improve cancer detection, but such strategies introduce additional trade-offs, including the risk of overdiagnosis, and do not, on their own, resolve the inferential challenge of disentangling risk from detectability \cite{4.2}. As a result, it remains unclear whether the risk attributed to dense tissue---compounded by the concern that masking may hide existing cancers---is large enough to justify the additional burden of early intervention. We aim to address this open question.

A further challenge is that screening histories are longitudinal and typically observed at irregular times. In routine care and screening programs, individuals rarely share the same assessment schedule, and visit timing can be related to underlying health status or clinical concern \cite{1.5,1.2}. Ignoring this irregular and potentially informative visit process can bias estimation of disease trajectories and risk relationships \cite{1.3}. Existing approaches for irregular longitudinal data in healthcare include but not limited to time-to-event models with time-dependent covariates, joint models for longitudinal and survival outcomes \cite{1.1}, mixed-effects models that account for informative visiting \cite{1.4}, and inverse-intensity weighting and doubly robust estimators tailored to irregular follow-up \cite{1.6}. Methodological reviews emphasize that no single approach is uniformly appropriate across all plausible visit mechanisms and recommend that the visit process be examined to guide model choice and analysis \cite{1.5}. Following this guidance, we adopt a regularization procedure tailored to the structure of our screening data.

In parallel, breast cancer modeling has seen rapid growth in machine learning and deep learning applications that prioritize predictive accuracy \cite{2.1,2.6,2.5}. Some studies report classification performance above 90\% \cite{2.2,2.4}. Large-scale evaluations also show that AI systems can surpass human judgement in image interpretation under some screening workflows and can reduce radiologist workload \cite{2.3}. Nevertheless, many of these models remain difficult to interpret clinically, often providing only binary predictions without considering inference or the longitudinal structure. Evidence suggests that changes in mammographic density over time have an impact on the breast cancer \cite{3.4}. Hence, methodologies that focus only on prediction and disregard longitudinal structure may be insufficient for understanding how risk evolves over time. In the current literature there is a need for interpretable approaches that account for longitudinal screening histories, represent disease progression, and address the detectability and risk challenge induced by dense tissue.  

In this paper, we address these gaps by proposing a progression--detection--risk framework for screening data with irregular follow-up, using mammographic density as the longitudinal response and cancer diagnosis as the event outcome. Our assumption is that density itself is not the primary
driver of cancer risk; rather, density reflects an underlying latent disease process that governs the observed density categories, and it is this latent process that is associated with cancer diagnosis. Under this assumption, the framework decouples risk from detectability and, crucially, allows us to
quantify how the estimated risk changes under different cancer detection probabilities. This quantification directly informs early-intervention decisions, which are currently made without
knowing how much masking causes us to underestimate the risk; measuring this gap reveals whether the underestimation is minor, or large enough to justify an earlier intervention in the subgroups most affected by masking.

This study also addresses two secondary questions: (i) how the relationship between mammographic density and latent risk differs across body mass index groups; and (ii) which patient characteristics influence progression patterns over follow-up. To test the main assumption and address these questions, we propose a multi-phase framework. First, we apply the regularization procedure. Second, to represent latent
progression, we use a latent Markov model \cite{5.2}. Third, we link the latent states to cancer diagnosis while incorporating detection probabilities, reflecting that diagnosis depends on both cancer development and detectability, and we propagate the uncertainty in the inferred latent states
throughout.

The remainder of the paper is organized as follows. Section~\ref{Data-desc} describes the data
collection process and the resulting data structure. Section~3 presents the proposed methodology for progression, detection, and risk. Section~\ref{sec-applied} applies this methodology to the dataset and reports the empirical results. Section~\ref{sec-discussion} concludes with a discussion of
implications and limitations.

\section{Breast cancer and mammographic density data}\label{Data-desc}

This study extends the analysis in \cite{3.4} and uses the same dataset. The data retain a longitudinal screening structure and include 616 patients from the Oncology Center at Ankara Etlik City Hospital. Patients were seen either for routine breast cancer screening or diagnosed with breast cancer during 2022--2024; mammographic density information, however, is retrieved retrospectively, and covers a longer, 6--10 years, of historical window and includes screenings recorded prior to this period. 

We included only patients with at least five mammographic density assessments before being diagnosed or having their routine screening. Consent form was collected from all the patients. The average interval between consecutive screenings was 14 months (SD 1.45) after excluding patients with gaps exceeding 24 months. Mammographic breast density was assessed using a four-level classification according to American College of Radiology's Breast Imaging Reporting and Data Systems (BIRADS): D, C, B, and A representing decreasing levels of risk \cite{3.9}. Category D is considered the highest risk (most dense), while category A corresponds to the lowest risk group. However, in our dataset, category A is never observed at any visit. All mammograms were interpreted by same four radiologists specializing in breast imaging at the hospital’s screening and diagnostic center. Mammograms obtained at external facilities were routinely re-read by the same team when images were available. Patients were excluded if external images could not be retrieved, if image quality was insufficient for reliable density assessment, or as mentioned if the spacing between any of the last five mammograms exceeded 24 months. 

Patient's characteristics were recorded at the last screening and included age (mean 53 years), menopausal status (47\% post-menopausal), family history of breast cancer (26\% positive), parity (median 2), and body mass index (BMI; mean 29). Patients with a breast cancer diagnosis were classified as cases (40\%), and those without a diagnosis as controls (60\%).The oldest age group (60+) had the lowest diagnosis proportion (37\%), followed by ages 33–49 (38\%) and 50–59 (41\%). Pre-menopausal (39\%) and post-menopausal (39\%) women shared the same level of diagnosis proportion. Patients with a positive family history had a higher diagnosis proportion (42\%) than those without (38\%). When parity was grouped into 0, 1–2, and 3+ births, diagnosis proportions were 37\%, 37\%, and 41\%, respectively. By BMI category, the diagnosis proportion was 37\% in the overweight group and 41\% in the obese group. 

\section{Methodology}\label{sec:method_lmm}

Our framework for decoupling risk and detection proceeds in two phases, preceded by a regularization step that resolves the irregular structure of the screening histories. In Phase~1, we model the longitudinal progression of risk through a latent process and, in doing so, separate the risk-relevant information from the observed mammographic density. In Phase~2, we quantify risk once it has been isolated from masking, and characterize the
trade-off between the degree of masking and the inferred risk. All three components are presented in Section~\ref{imput} (regularization), Section~\ref{sec:lmm} (Phase~1), and Section~\ref{sec:det_risk} (Phase~2).

The framework is implemented separately within body mass index (BMI) categories, so that the interpretation of mammographic density and its masking effect is allowed to vary across
BMI levels. This choice is motivated by the fact that density categories may not be directly comparable across BMI levels: the same observed classification can correspond to different
baseline risk profiles~\cite{5.6,5.7}. Moreover, BMI is closely related to adiposity, and hence to the relative amount of non-dense tissue, which can influence both the risk
attributed to a given density category and the degree of masking at screening~\cite{6.1,6.2,6.3}. Fitting a single model to the pooled cohort may therefore obscure these differences. For notational simplicity we omit the BMI-group index throughout; all model components are understood to be estimated separately within each BMI category.

\subsection{Irregular Missing Observations}\label{imput}

All 616 patients in our cohort are observed at irregular visit times, and only a few patients
share a common visit structure. We retain each patient's first five density assessments,
cutting the history at the fifth visit and tracking the four preceding visits. The
inter-visit gaps mostly reflect routine hospital scheduling,
and are therefore approximately regular up to minor variation. The fifth (cut-off) visit occurs between 2022 and 2024 for every patient. Hence, if we
trace back from the earliest first visit and forward to the latest fifth visit, all
individual histories fall within the window 2017--2024; the interval 2017--2024 therefore
covers every patient. The required extrapolation to the earliest first visit and the latest fifth visit spans
less than one year for each individual, indicating that it constitutes only a modest
extension of the observed follow-up. More fundamentally, the fact that a patient did not attend the hospital at a given
time point does not necessitate that the patient lacks a density category at that time, but
rather reflects that we simply do not observe it. Hence, in essence, the irregular-visit problem is only a missing-data problem within the window 2017--2024. 

As a simple illustration, suppose a patient visited the hospital on January~1 of each year from 2019 through 2023. Interpolation would concern the months between these January visits, while extrapolation would concern the period from January 2019 back to 2017 and from January 2023 forward to 2024. Thus, except for the five observed January assessments, every month on the 2017--2024 timeline would represent a missing density value for this patient. Every patient has the same general structure, with the only difference being the five months in which their density assessments were observed. For simplicity, the extrapolation in this illustrative example extends beyond one year; in the observed data, the required patient-level extrapolation is less than one year.

Motivated by this structural issue, we formalize the problem by discretizing the time frame into a monthly index
\(t = 1,\ldots,T\), where \(t=1\) corresponds to the earliest first visit (2017) and \(t=T\) to the latest fifth visit; in our data \(T = 80\). For each patient \(i\) and time point
\(t\), let \(\mathbf{x}_i^{(t)}\) denote the covariate vector and \(Y_i^{(t)}\) the density
response. For time points not aligned with an observed visit, \(Y_i^{(t)}\) is unobserved.
The covariates in our analysis are either time-invariant (e.g., family history, menopausal
status) or slowly varying (e.g., body mass index, parity), and are assumed unchanged and
carried forward across the grid. Missingness therefore concerns primarily the response
variable, the density category.

While imputation on this grid could in general be difficult, the structure of mammographic
density simplifies the problem substantially. Mammographic density declines over time through
the process of breast involution~\cite{determinantsMD2019,vanGils2013}, evolving predominantly
from denser toward less dense categories, \(D \rightarrow C \rightarrow B\), with reversals
occurring under rare cases such as hormone replacement therapy~\cite{HRTdensity}.
This near-monotone behavior is also seen in our cohort, where only two of 2464 observed
visit-to-visit transitions move in the reverse direction. We therefore exclude the patient exhibiting a reverse transition and assume monotonicity,
which allows most of the missingness to be resolved with a simple rule: for two consecutive
observed visits \(t_{ij}, t_{i,j+1} \in \mathcal{T}_i\) sharing the same category, every
intermediate time point takes that category,
\[
\text{if } Y_i^{(t_{ij})} = Y_i^{(t_{i,j+1})}, \quad\text{then}\quad
\tilde Y_i^{(t)} = Y_i^{(t_{ij})} \quad \text{for } t_{ij} < t < t_{i,j+1}.
\]
Likewise, time points before the first visit if the density is \(D\), and after the last
visit if the density is \(B\)---each less than one year of fill in our data---are extrapolated
by carrying the observed category backward or forward. The only remaining uncertain
regions are the gaps between two observed visits whose categories differ (patients moving
\(D\!\to\!C\) or \(C\!\to\!B\)), and the boundary segments with an intermediate endpoint (a
first visit of \(C\), or a last visit of \(C\) or \(D\)), which we fill probabilistically
using an estimated transition model.

We model the monthly density process as a monotone Markov chain on the ordered states
\(\{D, C, B\}\) with one-month transition matrix
\[
P =
\begin{pmatrix}
p_{DD} & p_{DC} & 0 \\
0 & p_{CC} & p_{CB} \\
0 & 0 & 1
\end{pmatrix},
\qquad p_{DC}=1-p_{DD}, \quad p_{CB}=1-p_{CC},
\]
where the absorbing state \(B\) and the zero upper-triangle encode monotonicity. For an
observed interval in which patient \(i\) moves from category \(a\) to \(b\) over \(g\) months,
the likelihood contribution is \(\Pr(Y_{i,t+g}=b \mid Y_{i,t}=a) = (P^{g})_{ab}\), and \(P\)
is estimated by maximizing
\[
\sum_i \sum_r \log\!\big\{ (P^{g_{ir}})_{a_{ir},\,b_{ir}} \big\},
\]
over all observed intervals after excluding the two reverse transitions.

Given the fitted matrix, missing gaps are
sampled from the Markov bridge that conditions on both observed endpoints,
\[
\Pr\!\big(Y_i^{(t+1)}=v \mid Y_i^{(t)}=u,\, Y_i^{(t+r)}=b\big)
=
\frac{P_{uv}\,(P^{\,r-1})_{vb}}{(P^{\,r})_{ub}},
\]
which preserves the observed endpoint categories, and
samples only monotone paths. This produces a complete, regularly indexed response sequence
\[
\tilde{\mathbf{Y}}_i = \big(\tilde Y_i^{(1)}, \ldots, \tilde Y_i^{(T)}\big),
\]
used as the observed sequence in the latent Markov model of Section~\ref{sec:lmm}.

We repeat the imputation \(R=3\) times, producing three completed versions of the 616-patient cohort. The full two-phase analysis is carried out on each version, and the resulting estimates are averaged within each bootstrap replicate as described in Section~\ref{sec:beta_inference}. For Phase~1 tables and figures that require a single completed cohort, we report one of the three versions selected at random. We emphasize that this procedure exploits the monotonicity of mammographic density and the short, regular follow-up of our cohort, and may not generalize to settings with rapidly varying outcomes, noisier measurements, or more irregular and prolonged follow-up.

\subsection{Phase 1: Latent Markov Model of Density Progression}\label{sec:lmm}

Our central assumption is that mammographic density itself does not carry breast cancer risk;
rather, the risk is carried by an unobserved progression state, of which density is only an
observed manifestation. Under this view, once the latent-state information is removed from the
density, what remains reflects only its masking aspect---the effect of dense tissue on
detectability. This separation is what allows us to decouple risk from masking, and we evaluate
it empirically in the Phase 2 model (Section~\ref{sec:det_risk}), where adding the final density
as a covariate contributes no significant information beyond the latent-state history.

\begin{figure}[h!]
    \centering
    \includegraphics[scale = 0.5]{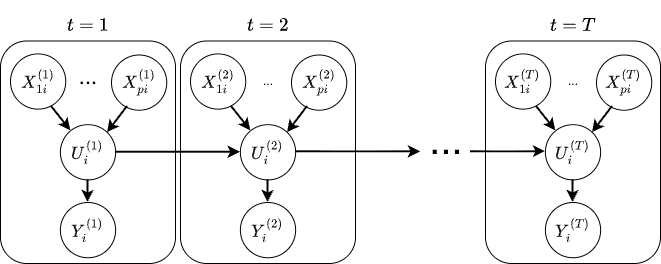}
    \caption{Directed acyclic graph (DAG) illustrating the hypothesized directional relationships
    between patient-specific covariates, the unobserved latent progression state, and the observed
    response variable.}
    \label{fig:DAG}
\end{figure}

To formalize this, we model each patient's regularized density sequence as arising from an
unobserved latent state that evolves over time according to a discrete-time Markov
process~\cite{5.2}. Having resolved the irregular observation times in Section~\ref{imput}, we
work on a common time index \(t = 1,\ldots,T\) across all subjects. The resulting structure is summarized by the graph in Figure~\ref{fig:DAG}.

For variable notations, we adopt the convention in Bartolucci et al.\ (2012)\cite{5.2}, with the exception that the time index is expressed as a superscript instead of a subscript. Let \(n\) denote
the number of patients and \(T\) the common number of time points after imputation
(Section~\ref{imput}). For each patient \(i \in \{1,\ldots,n\}\) and time point \(t \in
\{1,\ldots,T\}\), let
\[
\tilde{\mathbf{Y}}_i = (\tilde Y_i^{(1)}, \tilde Y_i^{(2)}, \ldots, \tilde Y_i^{(T)})
\]
denote the regularized univariate response sequence used for model fitting, and let
\[
\mathbf{U}_i = (U_i^{(1)}, U_i^{(2)}, \ldots, U_i^{(T)})
\]
denote the corresponding sequence of unobserved latent states, where \(U_i^{(t)} \in
\{1,\ldots,K\}\) and \(\tilde Y_i^{(t)} \in \{1,\ldots,c\}\), with \(K\) the number of latent
states and \(c\) the number of response categories. At each time point \(t\), the vector of \(p\)
covariates is \(\mathbf{x}_i^{(t)} = (x_{i1}^{(t)}, x_{i2}^{(t)}, \ldots, x_{ip}^{(t)})\), and the full covariate
matrix for patient \(i\) is \(\mathbf{X}_i = (\mathbf{x}_i^{(1)}, \mathbf{x}_i^{(2)}, \ldots,
\mathbf{x}_i^{(T)})\). The latent Markov model with this covariate structure consists of three
main components:

\begin{enumerate}
    \item \textbf{Initial state distribution:} The probability that patient \(i\) is in latent state \(u\) at baseline (\(t=1\)):
    \begin{equation}
        \pi_{i,u} = P(U_i^{(1)} = u), \quad u = 1, \ldots, K.
    \end{equation}\label{eq_1}
    
    \item \textbf{Transition probabilities:} The probability of transitioning from latent state \(u\) at time \(t{-}1\) to latent state \(v\) at time \(t\):
    \begin{equation}
        \pi^{(t)}_{i,uv} = P(U_i^{(t)} = v \mid U_i^{(t-1)} = u), \quad u,v = 1, \ldots, K;\ t = 2, \ldots, T.
    \end{equation}\label{eq_2}
    
    \item \textbf{Emission probabilities:} The probability of observing response \(y\) at time \(t\), given latent state \(u\):
    \begin{equation}
        \phi_{y\mid u} = P(\tilde Y_i^{(t)} = y \mid U_i^{(t)} = u), \quad y = 1, \ldots, c;\ u = 1, \ldots, K;\ t = 1, \ldots, T.
    \end{equation}\label{eq_3}
\end{enumerate}

Using Equation~\ref{eq_1}, the initial state distribution can be parameterized with a multinomial logistic model:
\begin{equation}
\log\left( \frac{\pi_{i,u}}{\pi_{i,1}} \right) = \beta_{0u} + \mathbf{x}_i^{(1)\top} \boldsymbol{\beta}_{1u}, \quad u = 2, \ldots, K.
\end{equation}\label{eq_4}

Similarly, through Equation~\ref{eq_2}, transition probabilities are modeled as:
\begin{equation}
\log\left( \frac{\pi^{(t)}_{i,uv}}{\pi^{(t)}_{i,uu}} \right) = \gamma_{0uv} + \mathbf{x}_i^{(t)\top} \boldsymbol{\gamma}_{1uv}, \quad u \ne v,\ t = 2, \ldots, T
\end{equation}\label{eq_5}

Parameter estimation is carried out via the Expectation-Maximization (EM) algorithm by maximizing the conditional log-likelihood of the regularized data. In particular, the complete-regularized data log-likelihood takes the form
\begin{align}
\ell_c(\boldsymbol{\theta}) =\ 
&\sum_{i=1}^n \sum_{t=1}^T \sum_{u=1}^K I(U_i^{(t)} = u)\, \log \phi_{\tilde y_i^{(t)}\mid u} \nonumber\\
&+ \sum_{i=1}^n \sum_{u=1}^K I(U_i^{(1)} = u)\, \log \pi_{i,u} \nonumber\\
&+ \sum_{i=1}^n \sum_{t=2}^T \sum_{u=1}^K \sum_{v=1}^K I(U_i^{(t-1)} = u,\, U_i^{(t)} = v)\, \log \pi^{(t)}_{i,uv}.
\end{align}\label{eq6}

Convergence is assessed based on stabilization of the regularized data log-likelihood. Once convergence is achieved for each BMI category when fitting the LMM using \texttt{LMest}\cite{5.3} package in RStudio, we conclude Phase~1 of our analysis.

\subsection{Phase 2: Decoupling detection and risk}
\label{sec:det_risk}
From Phase~1, we obtain summaries of each patient's latent progression history \(\mathbf{U}_i\),
which characterize how risk accumulates over follow-up. In this phase we incorporate cancer
detectability and quantify how the odds of a diagnosis change with one additional month spent in
each latent state, relative to a reference category. Let \(C_i \in \{0,1\}\) denote whether
patient \(i\) has developed cancer by the end of follow-up, and let \(D_i \in \{0,1\}\) denote
whether an existing cancer is detected at the final screening visit. We observe only the binary
diagnosis outcome
\[
Z_i \;=\; C_i\,D_i,
\]
so that a positive diagnosis at the final visit requires both cancer development and successful
detection. Because \(C_i\) and \(D_i\) are never observed separately, the decoupling of risk from
detectability is achieved through two structural assumptions with empirical justification, stated below together.

Define the latent-state occupancy counts
\[
n_{i}^j = \sum_{t=1}^T I\!\bigl(U_i^{(t)} = j\bigr), \qquad j = 1,\ldots,K,
\]
and let \(c_i = \tilde Y_i^{(T)}\) denote the density category at the final visit. Let
\(p_B, p_C, p_D \in (0,1]\) denote the detection probabilities for density categories
\(B, C, D\), with the clinically motivated ordering
\[
p_D < p_C < p_B,
\]
reflecting that denser tissue lowers mammographic sensitivity.

\begin{assumption}[Risk carried by latent history]\label{ass:risk}
Conditional on the cumulative latent-state occupancy and covariates, the observed density
trajectory carries no additional information about cancer development:
\[
\Pr\!\left(C_i = 1 \mid \{n_{i}^j\}, \tilde Y_i^{(1:T)}, \mathbf{x}_i^{(T)}\right)
=
\Pr\!\left(C_i = 1 \mid \{n_{i}^j\}, \mathbf{x}_i^{(T)}\right).
\]
\end{assumption}

Assumption~\ref{ass:risk} formalizes the idea that the latent-state occupancy summaries capture
the risk-bearing component of the longitudinal density process, so that the observed trajectory
adds nothing further once this history is included. We model the resulting cancer-development
probability through logistic regression,
\[
\Pr\!\left(C_i = 1 \mid \{n_{i}^j\}, \mathbf{x}_i^{(T)}\right)
=
\operatorname{logit}^{-1}\!\left(\beta_0 + \sum_{j=2}^K \beta_{n^j}\, n_{i}^j
+ \boldsymbol{\beta}^\top \mathbf{x}_i^{(T)}\right),
\]
with latent state \(j=1\) as the reference category. 

\begin{assumption}[Detectability carried by final density]\label{ass:detect}
Conditional on a cancer being present, detection at the final visit depends on the density
trajectory only through the final observed category,
\[
\Pr\!\left(D_i = 1 \mid C_i = 1, \{n_{i}^j\}, \tilde Y_i^{(1:T)}, \mathbf{x}_i^{(T)}\right)
= \Pr\!\left(D_i = 1 \mid C_i = 1, \tilde Y_i^{(T)} = c_i\right)
= p_{c_i},
\]
and detection is impossible in the absence of cancer, \(\Pr(D_i = 1 \mid C_i = 0, \cdot) = 0\).
\end{assumption}

Since \(Z_i = C_i D_i\), a positive diagnosis requires \(C_i = 1\) and \(D_i = 1\); applying the
product rule and then Assumptions~\ref{ass:risk} and~\ref{ass:detect} gives
\[
\begin{aligned}
\Pr\!\left(Z_i = 1 \mid c_i, \{n_{i}^j\}, \mathbf{x}_i^{(T)}\right)
&= \Pr\!\left(D_i = 1 \mid C_i = 1, c_i\right)\,
   \Pr\!\left(C_i = 1 \mid \{n_{i}^j\}, c_i, \mathbf{x}_i^{(T)}\right)\\[2pt]
&= p_{c_i}\,
   \Pr\!\left(C_i = 1 \mid \{n_{i}^j\}, \mathbf{x}_i^{(T)}\right),
\end{aligned}
\]
so that
\begin{equation}
\Pr\!\left(Z_i = 1 \mid \tilde Y_i^{(T)} = c_i, \{n_{i}^j\}, \mathbf{x}_i^{(T)}\right)
=
p_{c_i}\;
\operatorname{logit}^{-1}\!\left(\beta_0 + \sum_{j=2}^K \beta_{n^j}\, n_{i}^j
+ \boldsymbol{\beta}^\top \mathbf{x}_i^{(T)}\right).
\label{eq:det_risk_model_laststate}
\end{equation}
This factorization is the formal sense in which the model decouples risk from masking: the
logistic term carries the risk of cancer development through the latent history, while \(p_{c_i}\)
carries detectability through the final density alone. 


\begin{justification}
Assumption~\ref{ass:risk} states that, given the latent-state occupancy, the final density
adds no further information about cancer development. Because \(C_i\) and \(D_i\) are not observed separately, we assess this indirectly by augmenting the risk component of
\eqref{eq:det_risk_model_laststate} with the final density category \(c_i=\tilde Y_i^{(T)}\), entered through coefficients
\(\boldsymbol{\delta}=(\delta_B,\delta_C,\delta_D)^\top\) with
\(\delta_B+\delta_C+\delta_D=0\):
\[
\Pr\!\left(Z_i=1\mid c_i,\{n_i^j\},\mathbf{x}_i^{(T)}\right)
=
p_{c_i}\;
\operatorname{logit}^{-1}\!\left(\beta_0+\sum_{j=2}^K\beta_{n^j}n_i^j
+\boldsymbol{\beta}^\top\mathbf{x}_i^{(T)}
+\boldsymbol{\delta}^\top\mathbf d_i\right),
\]
where \(\mathbf d_i\) is the indicator of patient \(i\)'s final density category. Under this structure, the density block contributes nothing beyond the latent-state history, giving
\[
H_0:\ \delta_B=\delta_C=\delta_D=0.
\]
We assess \(H_0\) using an efficient score test evaluated under the null model after averaging over the random density completions and posterior latent paths described in Section~\ref{sec:beta_inference}. A non-significant result fails to provide evidence against Assumption~\ref{ass:risk}; the test statistics and \(p\)-values are reported in Appendix~\ref{tab:appendix2}.
\end{justification}

With the decoupling established and Assumption~\ref{ass:risk} supported, estimation focuses on the
latent-state risk coefficients \(\beta_{n^j}\), which are the main parameters of interest. We
report each on the odds-ratio scale: the odds ratio for one additional month spent in latent
state \(j\), relative to the reference state, is
\[
\text{OR}_j = \exp(\beta_{n^j}),
\]
holding other latent exposures and covariates fixed. Finally, because \(p_B, p_C, p_D\) are
not identifiable from the observed diagnoses alone, we do not estimate them; instead we treat them
as sensitivity parameters and refit \eqref{eq:det_risk_model_laststate} over clinically admissible
values satisfying \(p_D < p_C < p_B\), assessing how the latent-state odds ratios depend on
the assumed detectability across density categories.

\subsubsection{Uncertainty Quantification}
\label{sec:beta_inference}

The occupancy counts \(\{n_i^j\}\) entering \eqref{eq:det_risk_model_laststate} are not observed;
they are derived from randomly completed density histories and an estimated latent process. Quantifying the uncertainty of \(\widehat{\boldsymbol\beta}\) therefore requires propagating three sources of variability: (i) uncertainty arising from the stochastic completion of the unobserved density histories; (ii) uncertainty in the latent state trajectories, since different posterior draws of the latent paths produce different occupancy counts and, consequently, different estimates; and (iii) uncertainty in the latent Markov parameters that are themselves estimates. We address the
first two by random density completion and posterior path sampling and the latter by subject-level bootstrapping.

For completion \(r=1,\ldots,R\), we generate a completed density history
\(\widetilde{\mathbf Y}^{(r)}\), fit the latent Markov model to obtain
\(\widehat{\boldsymbol\theta}^{(r)}\), and draw \(M\) full latent trajectories from the
posterior of the state path,
\[
\mathbf U_i^{(r,m)} \sim
\Pr\!\left(U_i^{(1:T)} \mid \widetilde{\mathbf Y}_i^{(r)}, \mathbf X_i;
\widehat{\boldsymbol\theta}^{(r)}\right),
\qquad r=1,\ldots,R,\quad m=1,\ldots,M.
\]
For each draw we recompute the occupancy counts
\(n_i^{j,(r,m)}=\sum_t I(U_i^{(t,r,m)}=j)\) and refit
\eqref{eq:det_risk_model_laststate}, yielding an estimate
\(\widehat\beta^{(r,m)}\) for each coefficient. To propagate uncertainty in
\(\widehat{\boldsymbol\theta}\), this is wrapped in a bootstrap that resamples subjects with
replacement: for replicate \(b=1,\ldots,B\), \(R\) completed density histories are generated,
the latent Markov model is refit on each completed resampled dataset to obtain
\(\widehat{\boldsymbol\theta}^{(b,r)}\), \(M\) posterior paths are drawn under each
\(\widehat{\boldsymbol\theta}^{(b,r)}\), and Phase~2 is refit for each, producing
\(\widehat\beta^{(b,r,m)}\).

For a given coefficient, we first average over the \(M\) paths within each completed bootstrap
dataset and then over the \(R\) completed datasets within each bootstrap replicate,
\[
\bar\beta^{(b,r)}
=
\frac{1}{M}\sum_{m=1}^{M}\widehat\beta^{(b,r,m)},
\qquad
\bar\beta^{(b)}
=
\frac{1}{R}\sum_{r=1}^{R}\bar\beta^{(b,r)}.
\]
The point estimate is obtained by applying the same averaging to the original sample,
\[
\bar\beta^{(r)}
=
\frac{1}{M}\sum_{m=1}^{M}\widehat\beta^{(r,m)},
\qquad
\widehat\beta
=
\frac{1}{R}\sum_{r=1}^{R}\bar\beta^{(r)}.
\]
The empirical bootstrap variance is
\[
\widehat V_{\mathrm{boot}}
=
\frac{1}{B-1}\sum_{b=1}^{B}
\left(\bar\beta^{(b)}-\bar\beta_{\mathrm{boot}}\right)^2,
\qquad
\bar\beta_{\mathrm{boot}}
=
\frac{1}{B}\sum_{b=1}^{B}\bar\beta^{(b)}.
\]

Let \(q_{\alpha}^{*}\) denote the empirical \(\alpha\)-quantile of
\(\{\bar\beta^{(1)},\ldots,\bar\beta^{(B)}\}\). The point estimate
\(\widehat\beta\) and the bootstrap distribution yield the reported odds ratio and confidence
interval
\[
\mathrm{OR}=\exp(\widehat\beta),
\qquad
\left[\exp\!\left(q_{0.025}^{*}\right),\;
      \exp\!\left(q_{0.975}^{*}\right)\right].
\]

The same procedure is used to test the Assumption~\ref{ass:risk} indirectly. The null hypothesis is
\[
H_0:\ \delta_B=\delta_C=\delta_D=0,
\]
which has two degrees of freedom under the sum-to-zero constraint. The block is evaluated with an efficient score
test fitted under \(H_0\). For each patient, the null risk probability is first averaged over the \(R\) completed density histories and \(M\) posterior paths per completion. Let
\(\mathbf U_{\delta\cdot\beta}\) denote the resulting density score after adjustment for the
Phase~2 nuisance coefficients and let \(\mathbf I_{\delta\cdot\beta}\) denote its efficient
information. The density-block statistic is
\[
S_{\mathrm{density}}
=
\mathbf U_{\delta\cdot\beta}^{\top}
\mathbf I_{\delta\cdot\beta}^{-1}
\mathbf U_{\delta\cdot\beta}
\;\sim\;\chi_2^2
\quad\text{under }H_0.
\]
Its finite-sample distribution is evaluated using \(G\) conditional parametric null simulations
of the diagnosis outcome. If \(S_{\mathrm{density}}^{*(g)}\) denotes the statistic from
simulation \(g\), the bootstrap-calibrated value is
\[
p_{\mathrm{boot}}
=
\frac{1+\sum_{g=1}^{G}I\!\left(S_{\mathrm{density}}^{*(g)}
\geq S_{\mathrm{density}}\right)}{G+1}.
\]
For the two separately fitted BMI groups, the combined omnibus statistic is the sum of the two
group-specific statistics and is compared with a \(\chi_4^2\) reference distribution. The results are reported in Table~\ref{tab:appendix2}. The reported analysis uses \(R=3\), \(M=25\), and \(B=200\), with \(G=1000\) conditional parametric null simulations for the final-density score test.

\section{Application of the Method}\label{sec-applied}

We applied the two phase model to the longitudinal dataset described in Section~\ref{Data-desc}. Age, menopausal status, family history of breast cancer, and parity were included as covariates due to their influence on breast cancer \cite{5.5}. Mammographic breast density was modeled as the categorical longitudinal response in Phase~1 (Section~\ref{sec:lmm}), and breast cancer diagnosis status was modeled as the binary outcome in Phase~2 (Section~\ref{sec:det_risk}). 

The remainder of this section is organized as follows. Section~\ref{sec:res_lmm} reports Phase~1 progression results from the latent Markov model, including model fit and state specification (Section~\ref{sec:res_lmm_fit}), latent-state interpretation and labeling (Section~\ref{sec:res_lmm_label}), descriptive covariate summaries by state (Section~\ref{sec:res_lmm_cov}), and state evolution over follow-up (Section~\ref{sec:res_lmm_trans}). Section~\ref{sec:res_detrisk} reports Phase~2 detection and risk results. 


\subsection{Phase~1: Disease Progression}\label{sec:res_lmm}

Phase~1 reports results from fitting the latent Markov model in Section~\ref{sec:lmm} to the regularized longitudinal data (Section~\ref{imput}). All Phase~1 analyses are conducted within BMI categories, as described in Section~\ref{sec:method_lmm}. In this section, we summarize (i) model fit and the specification of the latent-state structure (Section~\ref{sec:res_lmm_fit}), (ii) interpretation and labeling of the latent states (Section~\ref{sec:res_lmm_label}), (iii) descriptive statistics (Section~\ref{sec:res_lmm_cov}), and (iv) estimated state transitions over time (Section~\ref{sec:res_lmm_trans}).

\subsubsection{Model fit and state specification}\label{sec:res_lmm_fit}
Fitting an LMM requires specifying the number of latent states \(K\), which determines the
resolution at which longitudinal density trajectories are summarized. We selected \(K\) by
combining empirical model fit with a clinically motivated reference. Clinically, invasive breast
cancer is commonly described in four stages (I--IV)~\cite{stages}, which provides a natural
benchmark for the latent-state resolution of the progression model. Empirically, LMMs with
\(K \in \{1,\ldots,7\}\) were fitted and compared via information criteria, with \(K=7\) taken as
an upper bound since changes beyond this range were negligible. The log-likelihood, number of parameters, AIC, and BIC for the overweight and obese categories are reported in Appendix~\ref{tab:appendix3}, respectively. In both groups the \(K=4\) model achieved the second-lowest BIC. Given this strong empirical performance and its alignment with the
four-stage clinical framework, we fixed \(K=4\) for all subsequent analyses in both BMI categories.

To confirm that the findings are not an artifact of this choice, we additionally repeated the full analysis at \(K=3\) and \(K=5\) (the latter giving the lowest BIC) as a sensitivity check.
The conclusions are consistent across both specifications in direction and significance; details are reported in Appendix~\ref{tab:appendix4}. All models converged successfully in both groups.

\subsubsection{Latent state interpretation and labeling}\label{sec:res_lmm_label}

Although the LMM estimates distinct latent states, the state labels are not identifiable a priori and must be interpreted post hoc. We therefore label states using two complementary summaries from the fitted model: (i) emission probabilities, and (ii) the state directions in the estimated transition matrix. Let $\tilde Y_i^{(t)} \in \{B,C,D\}$ denote the regularized mammographic density category at visit $t$, and let $U_i^{(t)} \in \{1,2,3,4\}$ denote the corresponding latent state. Since in our dataset, category $A$ is never observed at any visit, we restrict the density-category state space to $\{B,C,D\}$ and drop $A$ throughout. The emission probabilities are
\[
\phi_{d\mid u} \;=\; P\!\bigl(\tilde Y_i^{(t)} = d \mid U_i^{(t)} = u\bigr),
\qquad d\in\{B,C,D\},\ u\in\{1,2,3,4\}.
\]
Table~\ref{tab:emission_probs} reports \(\hat\phi_{d\mid u}\). The fitted emission distributions demonstrates different allocations of latent states across density categories by BMI group. In the overweight group, one latent state concentrates on \(B\), one concentrates on \(C\), and two distinct latent states concentrate on \(D\). In contrast, in the obese group, one latent state concentrates on \(B\), two distinct latent states concentrate on \(C\), and one concentrates on \(D\). 

\begin{table}[h!]
\centering
\begin{tabular}{c|cccc|cccc}
\toprule
\multirow{2}{*}{} & \multicolumn{4}{c|}{Overweight} & \multicolumn{4}{c}{Obese} \\
\cmidrule{2-9}
Density & LR (State 1) & EmR (State 2) & EsR (State 3) & CR (State 4) & LR (State 1) & EmR (State 2) & EsR (State 3) & CR (State 4) \\
\midrule
B & 1.0000 & 0.0000 & 0.0000 & 0.0000 & 1.0000 & 0.0000 & 0.0000 & 0.0000 \\
C & 0.0000 & 1.0000 & 0.0000 & 0.0000 & 0.0000 & 1.000 & 0.9998 & 0.0000 \\
D & 0.0000 & 0.0000 & 1.0000 & 1.0000 & 0.0000 & 0.0000 & 0.0002 & 1.0000 \\
\bottomrule
\end{tabular}
\caption{Estimated emission probabilities $\hat\phi_{d\mid u}=\Pr(\tilde Y=d\mid U=u)$ for $d\in\{B,C,D\}$ and reported separately by BMI category. LR, EmR, EsR, and CR denote Low Risk, Emergent Risk, Established Risk, and Critical Risk state labels.}
\label{tab:emission_probs}
\end{table}

We then use the estimated transition probabilities to order the four states and assign risk labels. Writing
\[
\pi_{uv} \;=\; P\!\bigl(U_i^{(t)} = v \mid U_i^{(t-1)} = u\bigr),
\qquad u,v\in\{1,2,3,4\},
\]
Table~\ref{tabtab:st2} indicates that state changes occur predominantly in a single direction similar to the density behavior. Together with the emission patterns above, this ordering motivates labeling the four states as low risk, emergent risk, established risk, and critical risk. Under this labeling, transitions proceed from higher- to lower-risk states. This perspective also clarifies why our primary exposure of interest is cumulative time spent in each latent state, rather than the observed density category at any single visit: although many individuals tend to move toward lower-risk states and less dense categories over follow-up, current risk cannot be inferred from the contemporaneous category alone because patients are not independent of their history and may carry accumulated risk from earlier high-risk periods. 

\begin{table}[h!]
\centering
\begin{tabular}{l|cccc|cccc}
\toprule
\multirow{2}{*}{} & \multicolumn{4}{c|}{Overweight} & \multicolumn{4}{c}{Obese} \\
\cmidrule{1-9}
From $\backslash$ To & CR (State 4) & EsR (State 3) & EmR (State 2) & LR (State 1) & CR (State 4) & EsR (State 3) & EmR (State 2) & LR (State 1) \\
\midrule
CR (State 4) & 0.955 & 0.040 & 0.005 & 0.000 & 0.970 & 0.030 & 0.000 & 0.000 \\
EsR (State 3) & 0.012 & 0.907 & 0.081 & 0.000 & 0.000 & 0.959 & 0.041 & 0.000 \\
EmR (State 2) & 0.000 & 0.000 & 0.982 & 0.018 & 0.000 & 0.003 & 0.951 & 0.047 \\
LR (State 1) & 0.000 & 0.000 & 0.000 & 1.000 & 0.000 & 0.000 & 0.000 & 1.000 \\
\bottomrule
\end{tabular}
\caption{Estimated transition probabilities $\hat{\pi}_{uv}$ between labeled latent states, shown separately by BMI category. LR, EmR, EsR, and CR correspond to Low Risk, Emergent Risk, Established Risk, and Critical Risk state labels.}
\label{tabtab:st2}
\end{table}
Under the same labeling, the model reveals how mammographic density categories map onto latent risk states differently across BMI groups. In the overweight group, density \(D\) is resolved into two distinct latent states, whereas in the obese group it is density \(C\) that splits into two states. In other words, the density category that the model finds rich enough to warrant finer-grained latent structure shifts for different BMI groups. This is consistent with our expectation that, at higher BMI, a given amount of risk-bearing tissue presents at a lower observed density category: the heterogeneity the model attributes to density \(D\) in overweight patients appears instead within density \(C\) in obese patients. Density alone is therefore an incomplete risk indicator, since the latent-state composition underlying the same observed category differs between groups. The risk interpretation of the latent states is further supported by the remaining results.



\subsubsection{Summary statistics for inferred states}\label{sec:res_lmm_cov}

We examined how patient characteristics vary across the four labeled latent states and how these profiles change over follow-up. Tables~\ref{tab:summary_stats_ow}--\ref{tab:summary_stats_ob} report covariate summaries within each state at three points in the visit sequence (first, middle, last) for the overweight and obese BMI groups, respectively. In both BMI groups, the first-visit distributions show no patients in the low-risk state. Moreover, no patients have mammographic density recorded as \(B\) at the first visit. Consequently, first-visit comparisons are restricted to the emergent-, established-, and critical-risk states. In the overweight BMI group at the first visit, family history is markedly more common in the critical risk state than in the lower-risk states, occurring about three times as often (40\% versus 19\% and 13\% in established and emergent states). In the opposite direction, post-menopausal status is more common in the less risky state: patients in emergent risk are approximately twice as likely to be post-menopausal as those in critical risk (64\% versus 26\%).

Over follow-up, the distribution of patients across the latent risk states shifts toward the lower-risk states. By the last visit, 97\% of overweight patients are in the two least risky, low and emergent risk states. At the last visit, similar relationship in first visit is preserved: family history remains more frequent in the emergent risk state than in the low risk state (35\% versus 27\%), whereas post-menopausal status is less frequent (34\% versus 43\%).

\begin{table}[h!]
\centering
\renewcommand{\arraystretch}{1.2}
\setlength{\tabcolsep}{2pt}
\small
\begin{tabular}{l|cccc|cccc|cccc}
\toprule
\multirow{2}{*}{} & \multicolumn{4}{c|}{First visit} & \multicolumn{4}{c|}{Middle visit} & \multicolumn{4}{c}{Last visit} \\
\cmidrule{2-13}
Covariate & LR & EmR & EsR & CR & LR & EmR & EsR & CR & LR & EmR & EsR & CR \\
\midrule
Age & -- & 50.3 (5.7) & 39.3 (4.3) & 44.8 (5.3) & 53.0 (5.9) & 50.5 (6.2) & 44.4 (3.9) & 48.0 (5.9) & 54.1 (6.0) & 49.7 (5.5) & 49.0 & 46.0 (7.9) \\
BMI & -- & 28.4 (0.8) & 27.9 (1.0) & 28.0 (1.1) & 28.4 (0.8) & 28.2 (1.0) & 27.9 (1.0) & 27.9 (1.1) & 28.2 (0.9) & 27.9 (1.1) & 29.0 & 27.7 (1.2) \\
Positive family history & -- & 13\% & 19\% & 40\% & 14\% & 23\% & 38\% & 42\% & 27\% & 35\% & 0\% & 30\% \\
Parity & -- & 2.3 (1.3) & 1.4 (1.1) & 1.5 (1.0) & 2.2 (1.4) & 2.0 (1.2) & 1.3 (1.1) & 1.4 (1.0) & 2.1 (1.2) & 1.3 (1.0) & 2.0 & 1.2 (0.9) \\
Post-menopausal & -- & 64\% & 40\% & 26\% & 54\% & 43\% & 33\% & 26\% & 43\% & 34\% & 100\% & 0\% \\
Density B & -- & 0\% & 0\% & 0\% & 100\% & 0\% & 0\% & 0\% & 100\% & 0\% & 0\% & 0\% \\
Density C & -- & 100\% & 0\% & 0\% & 0\% & 100\% & 0\% & 0\% & 0\% & 100\% & 0\% & 0\% \\
Density D & -- & 0\% & 100\% & 100\% & 0\% & 0\% & 100\% & 100\% & 0\% & 0\% & 100\% & 100\% \\
\midrule
Size ($n$) & 0 & 105 & 38 & 229 & 28 & 188 & 84 & 72 & 225 & 136 & 1 & 10 \\
\bottomrule
\end{tabular}
\caption{Overweight BMI group: covariate summaries within each latent state at the first, middle, and last screening visit. Continuous variables are reported as mean (SD) and binary indicators as percentages.}
\label{tab:summary_stats_ow}
\end{table}

In the obese BMI group at the first visit, the same opposing patterns are observed across latent risk states. Family history is more frequent in the higher-risk state: patients assigned to critical risk report a family history approximately twice as often as those in established risk (29\% versus 13\%). On the other hand, post-menopausal status is more common in the less risky state: patients in established risk are about twice as likely to be post-menopausal as those in critical risk (76\% versus 37\%). Over follow-up, similar to overweight group, the state distribution shifts away from critical risk; by the last visit, all but three patients are in low risk or established risk. Within this distribution, the same ordering persists: family history remains twice as common in established risk as in low risk (32\% versus 16\%), whereas post-menopausal status is more prevalent in low risk than established risk (66\% versus 42\%). 

\begin{table}[h!]
\centering
\renewcommand{\arraystretch}{1.2}
\setlength{\tabcolsep}{2pt}
\small
\begin{tabular}{l|cccc|cccc|cccc}
\toprule
\multirow{2}{*}{} & \multicolumn{4}{c|}{First visit} & \multicolumn{4}{c|}{Middle visit} & \multicolumn{4}{c}{Last visit} \\
\cmidrule{2-13}
Covariate & LR & EmR & EsR & CR & LR & EmR & EsR & CR & LR & EmR & EsR & CR \\
\midrule
Age & -- & 48.8 (5.9) & 51.1 (5.3) & 45.8 (5.9) & 55.7 (5.2) & 49.3 (4.4) & 55.1 (6.0) & 48.4 (5.5) & 56.5 (5.8) & -- & 52.5 (6.7) & 48.7 (3.8) \\
BMI & -- & 33.5 (2.8) & 31.9 (1.9) & 31.0 (1.3) & 32.9 (2.5) & 31.7 (1.7) & 31.6 (1.7) & 30.7 (0.9) & 31.8 (1.9) & -- & 31.1 (1.6) & 31.1 (0.8) \\
Positive family history & -- & 13\% & 13\% & 29\% & 17\% & 18\% & 15\% & 32\% & 16\% & -- & 31\% & 33\% \\
Parity & -- & 4.8 (0.9) & 3.0 (1.2) & 2.0 (1.2) & 3.5 (1.4) & 2.7 (1.3) & 2.7 (1.4) & 2.0 (1.2) & 2.9 (1.3) & -- & 2.1 (1.4) & 1.0 (1.7) \\
Post-menopausal & -- & 73\% & 76\% & 37\% & 86\% & 49\% & 70\% & 44\% & 66\% & -- & 42\% & 33\% \\
Density B & -- & 0\% & 0\% & 0\% & 100\% & 0\% & 0\% & 0\% & 100\% & -- & 0\% & 0\% \\
Density C & -- & 100\% & 100\% & 0\% & 0\% & 100\% & 100\% & 0\% & 0\% & -- & 100\% & 0\% \\
Density D & -- & 0\% & 0\% & 100\% & 0\% & 0\% & 0\% & 100\% & 0\% & -- & 0\% & 100\% \\
\midrule
Size ($n$) & 0 & 15 & 135 & 94 & 36 & 87 & 80 & 41 & 193 & 0 & 48 & 3 \\
\bottomrule
\end{tabular}
\caption{Obese BMI group: covariate summaries within each latent state at the first, middle, and last screening visit. Continuous variables are reported as mean (SD) and binary indicators as percentages.}
\label{tab:summary_stats_ob}
\end{table}

Taken together, these patterns indicate that, in both BMI groups and across visits, the higher-risk latent states more often coincide with a stronger familial predisposition and a pre-menopausal profile. In the next section, we move beyond these descriptive summaries and report how the covariates relate to transitions between latent states using the transition model.

\subsubsection{Estimated state transitions parameters}\label{sec:res_lmm_trans}

Section~\ref{sec:res_lmm_label} indicates that estimated transitions occur predominantly from higher-risk states toward lower-risk states. In this section, we summarize which covariates influence the (i) the latent state at the first screening visit, and (ii) progression patterns across states over follow-up for the overweight and obese BMI groups. We begin with the initial latent state model at the first observed visit. Table~\ref{tab:initial_or_combined_k4} reports odds ratios from the multinomial logit specification for \(U_i^{(1)}\), with low risk as the reference state.In the overweight group, the initial state is closely aligned with menopausal status: post-menopausal patients are far more likely to begin in the established risk state than in low risk. The association is so strong that the corresponding odds ratio is effectively unbounded, reflecting near-complete separation. Among pre-menopausal patients, the baseline state is instead graded by age: older patients have substantially higher odds of starting in the critical risk state (OR \(= 2.78^{***}\)) and the emergent risk state (OR \(= 2.40^{***}\)) relative to low risk.
In summary, an overweight patient who is post-menopausal is most likely in the established risk state at their first visit to the hospital; if still pre-menopausal, the baseline state is graded by age, with older patients tending to be in critical or emergent risk and younger patients in low risk.

In the obese group, the baseline ordering differs. Post-menopausal status aligns with the emergent risk state rather than the established risk state; as in the overweight group, this association is effectively unbounded. Among pre-menopausal patients, baseline separation is captured primarily by age and parity: higher age and lower parity increase the odds of starting in the critical risk state (OR \(= 2.83^{***}\) for age, \(0.01^{*}\) for parity) and the established risk state (OR \(= 3.09^{***}\) for age, \(0.01^{**}\) for parity) relative to low risk. In summary, an obese patient who is post-menopausal is most likely in the emergent risk state at
their first visit to the hospital; if still pre-menopausal, the baseline state is graded by age and parity, with older, lower-parity patients tending to be in critical or established risk and younger, higher-parity patients in low risk.

\begin{table}[h!]
\centering
\begin{tabular}{l|lll|lll}
\toprule
\multirow{2}{*}{} & \multicolumn{3}{c|}{Overweight} & \multicolumn{3}{c}{Obese} \\
\cmidrule{2-7}
Covariate & EmR & EsR & CR & EmR & EsR & CR\\
\midrule
Age & $2.866^{***}$ & $0.244^{***}$ & $2.470^{***}$ & $0.246^{***}$ & $3.052^{***}$ & $2.801^{***}$ \\
Post-menopausal & $0.002^{***}$ & $77305.195^{***}$ & $0.002^{***}$ & $3263.667^{**}$ & $0.007^{**}$ & $0.004^{***}$ \\
Positive family history & $26.179^{***}$ & $0.002^{***}$ & $71.944^{***}$ & $0.860$ & $1.393$ & $2.329$ \\
Parity & $0.859$ & $1.256$ & $0.623$ & $3655.935^{**}$ & $0.015^{**}$ & $0.010^{***}$ \\
\bottomrule
\end{tabular}
\caption{Initial-state odds ratios from the multinomial logit model for \(U_i^{(1)}\), with low risk (LR) as the reference state, reported separately for overweight and obese BMI groups. Significance codes: \(^{*}p<0.1\), \(^{**}p<0.05\), \(^{***}p<0.001\).}
\label{tab:initial_or_combined_k4}
\end{table}
Following baseline, in Table~\ref{tab:transition_params_k4_updated}, we summarize which covariates relate to movement from more risky toward less risky latent states over follow-up, reported separately for the overweight and obese BMI groups. As shown in Section~\ref{sec:res_lmm_label}, the fitted transition structure permits mostly movement toward less risky states (i.e., transitions to more risky states have mostly estimated probability 0). Accordingly, Table~\ref{tab:transition_params_k4_updated} reports covariate associations for downward transitions.

From the critical risk state, patients may remain there or move to any less risky state. With increasing age, a single-step transition directly to low risk becomes less likely in both groups (overweight: OR $=0.861^{***}$; obese: OR $=0.848^{***}$), while age raises the odds of an intermediate downward step---into emergent risk for the overweight group (OR $=2.034^{***}$) and into established risk for the obese group (OR $=1.084^{***}$).

Menopausal status differs across BMI groups. In the overweight group, pre-menopausal patients (vs.\ post-menopausal) have higher odds of moving toward less risky states (e.g.\ critical-to-low OR $=0.756^{***}$, established-to-low OR $=0.619^{***}$), whereas in the obese group the direction is largely reversed (critical-to-low OR $=1.585^{***}$, established-to-low OR $=3.293^{***}$).

A family history of breast cancer (vs.\ none) is associated with lower odds of de-escalating from the established risk state in both groups (overweight OR $=0.169^{***}$; obese OR $=0.045^{***}$). Parity shows opposing associations across groups: higher parity raises the odds of de-escalating to low risk in the overweight group (e.g.\ established-to-low OR $=1.955^{***}$) but lowers them in the obese group (OR $=0.594^{***}$).

In short, a younger patient without a family history de-escalates toward low risk more readily regardless of BMI group. De-escalation slows, however, for post-menopausal patients with lower parity in the overweight group, and for pre-menopausal patients with higher parity in the obese group.

\begin{table}[h!]
\centering
\begin{tabular}{l|l|lll|lll}
\toprule
\multirow{1}{*}{} & \multirow{1}{*}{} & \multicolumn{3}{c|}{Overweight} & \multicolumn{3}{c}{Obese} \\
\cmidrule{1-8}
Current State & Covariate & $\rightarrow$ EsR & $\rightarrow$ EmR & $\rightarrow$ LR & $\rightarrow$ EsR & $\rightarrow$ EmR & $\rightarrow$ LR \\
\midrule
\multirow{4}{*}{CR} & Age & $0.965$ & $1.598^{***}$ & $0.854^{***}$ & $1.094^{***}$ & $0.621^{***}$ & $0.846^{***}$ \\
 & Post-menopausal & $1.040$ & $23.974$ & $0.379^{***}$ & $0.432^{**}$ & $122.541^{***}$ & $1.485^{***}$ \\
 & Positive family history & $1.207$ & $0.343$ & $0.494^{***}$ & $1.221$ & $0.027^{***}$ & $0.541^{***}$ \\
 & Parity & $1.154$ & $0.388$ & $1.451^{***}$ & $0.980$ & $5.190^{***}$ & $1.680^{***}$ \\
\midrule
\multirow{4}{*}{EsR} & Age & -- & $1.219^{***}$ & $1.015^{***}$ & -- & $0.976$ & $0.896^{***}$ \\
 & Post-menopausal & -- & $0.244^{***}$ & $0.079^{***}$ & -- & $0.985$ & $0.176^{***}$ \\
 & Positive family history & -- & $1.042$ & $0.101^{***}$ & -- & $1.006$ & $26.853^{***}$ \\
 & Parity & -- & $0.947$ & $1.687^{***}$ & -- & $1.049$ & $0.710^{***}$ \\
\midrule
\multirow{4}{*}{EmR} & Age & -- & -- & $1.145^{***}$ & -- & -- & $1.173^{***}$ \\
 & Post-menopausal & -- & -- & $0.195^{***}$ & -- & -- & $0.282^{**}$ \\
 & Positive family history & -- & -- & $1.236$ & -- & -- & $0.903$ \\
 & Parity & -- & -- & $1.142^{**}$ & -- & -- & $0.911$ \\
\bottomrule
\end{tabular}
\caption{Transition odds ratios from the latent Markov model, reported separately by current state. For each current state, odds ratios compare transitioning to the indicated lower-risk state versus remaining in the current state (reference). Non-applicable transitions are shown as ``--''. Statistical significance: $^{*}p<0.1$, $^{**}p<0.05$, $^{***}p<0.001$.}
\label{tab:transition_params_k4_updated}
\end{table}

\subsubsection{State transitions over time}\label{sec:res_lmm_trans}

\begin{figure}[h!]
\centering

\begin{subfigure}[b]{0.48\linewidth}
\centering
\includegraphics[scale = 0.22]{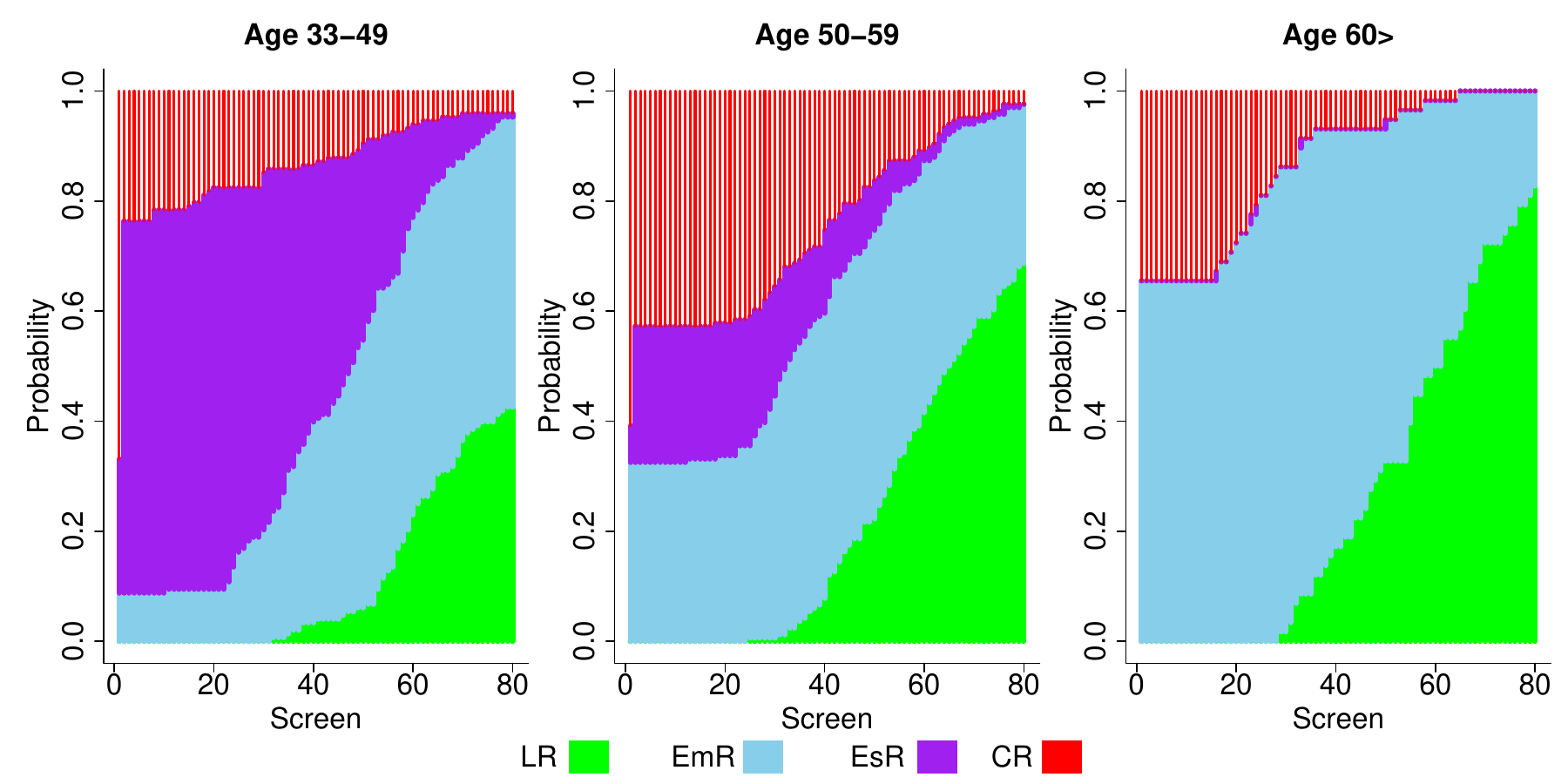}
\caption*{\footnotesize{(a) Age: Overweight}}
\label{fig:ow_age}
\end{subfigure}
\hfill
\begin{subfigure}[b]{0.48\linewidth}
\centering
\includegraphics[scale = 0.18]{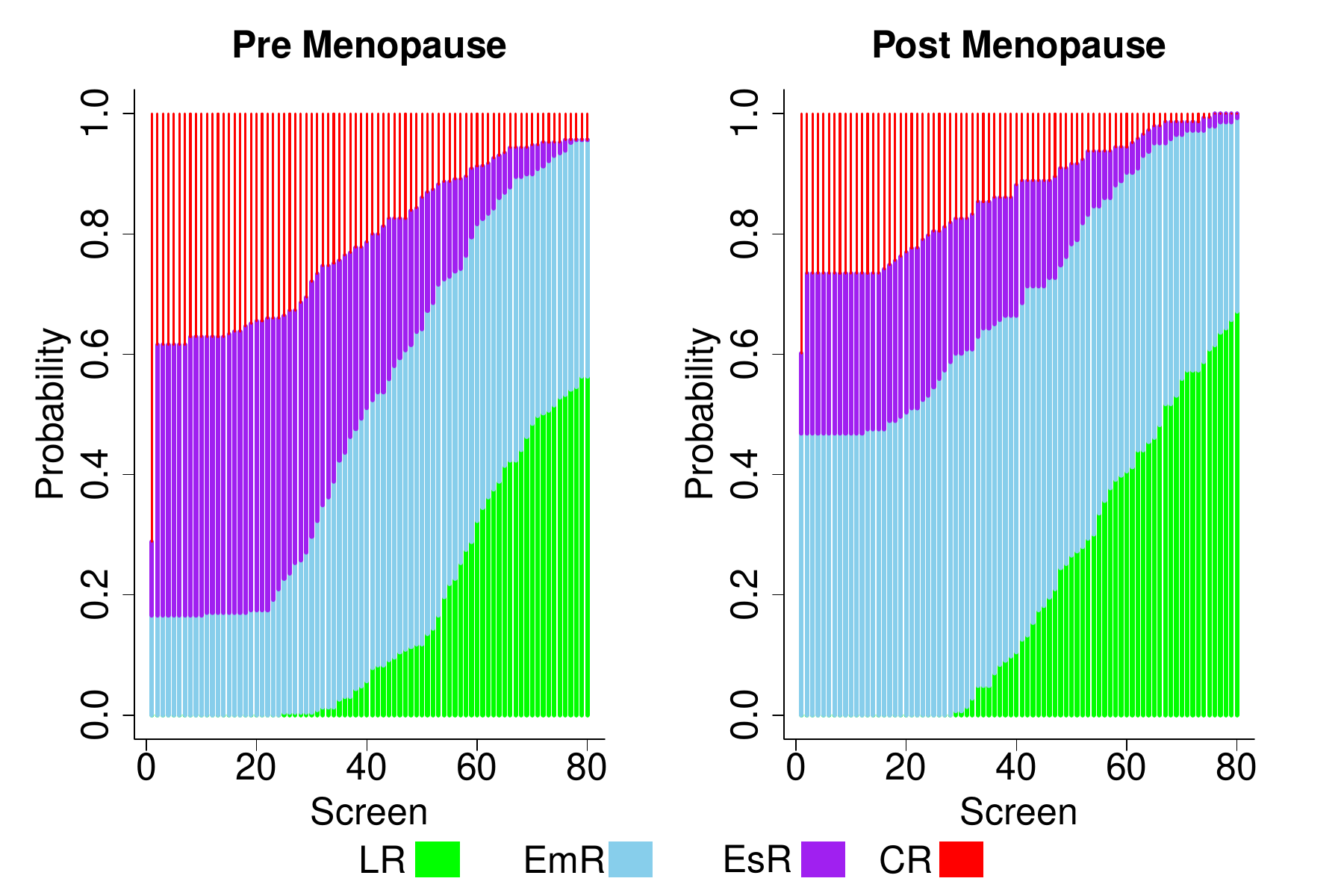}
\caption*{\footnotesize{(c) Menopause: Overweight}}
\label{fig:ow_meno}
\end{subfigure}

\vspace{0.8em}

\begin{subfigure}[b]{0.48\linewidth}
\centering
\includegraphics[scale = 0.22]{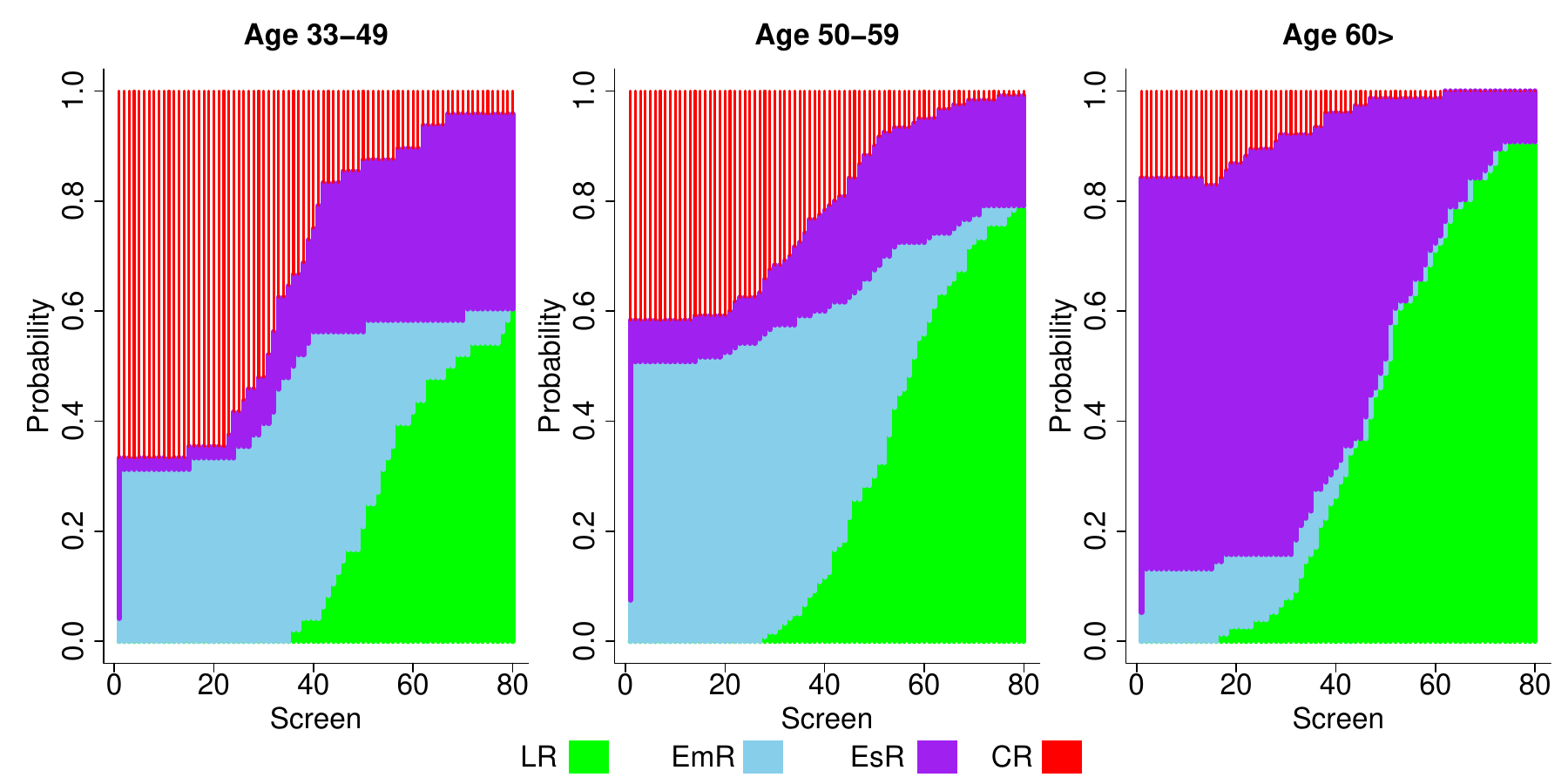}
\caption*{\footnotesize{(b) Age: Obese}}
\label{fig:ob_age}
\end{subfigure}
\hfill
\begin{subfigure}[b]{0.48\linewidth}
\centering
\includegraphics[scale = 0.18]{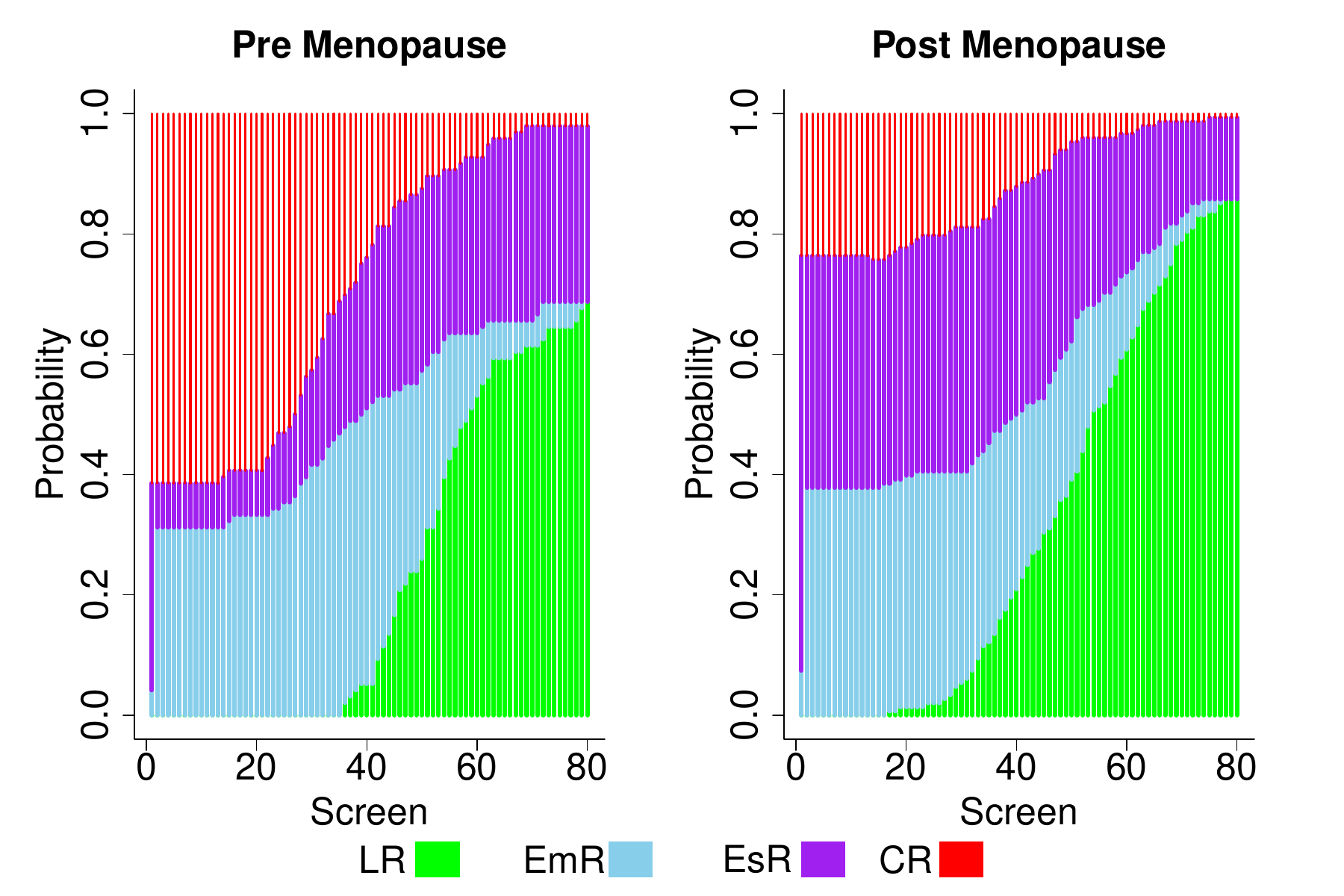}
\caption*{\footnotesize{(d) Menopause: Obese}}
\label{fig:ob_meno}
\end{subfigure}

\caption{Estimated latent-state proportions over screening visits, shown separately for the overweight and obese BMI groups, by age group (33--49, 50--59, 60+; left column) and menopausal status (pre- vs.\ post-menopausal; right column).}
\label{fig:age_meno_profiles}
\end{figure}

\begin{figure}[h!]
\centering

\begin{subfigure}[b]{0.48\textwidth}
    \centering
    \includegraphics[scale = 0.18]{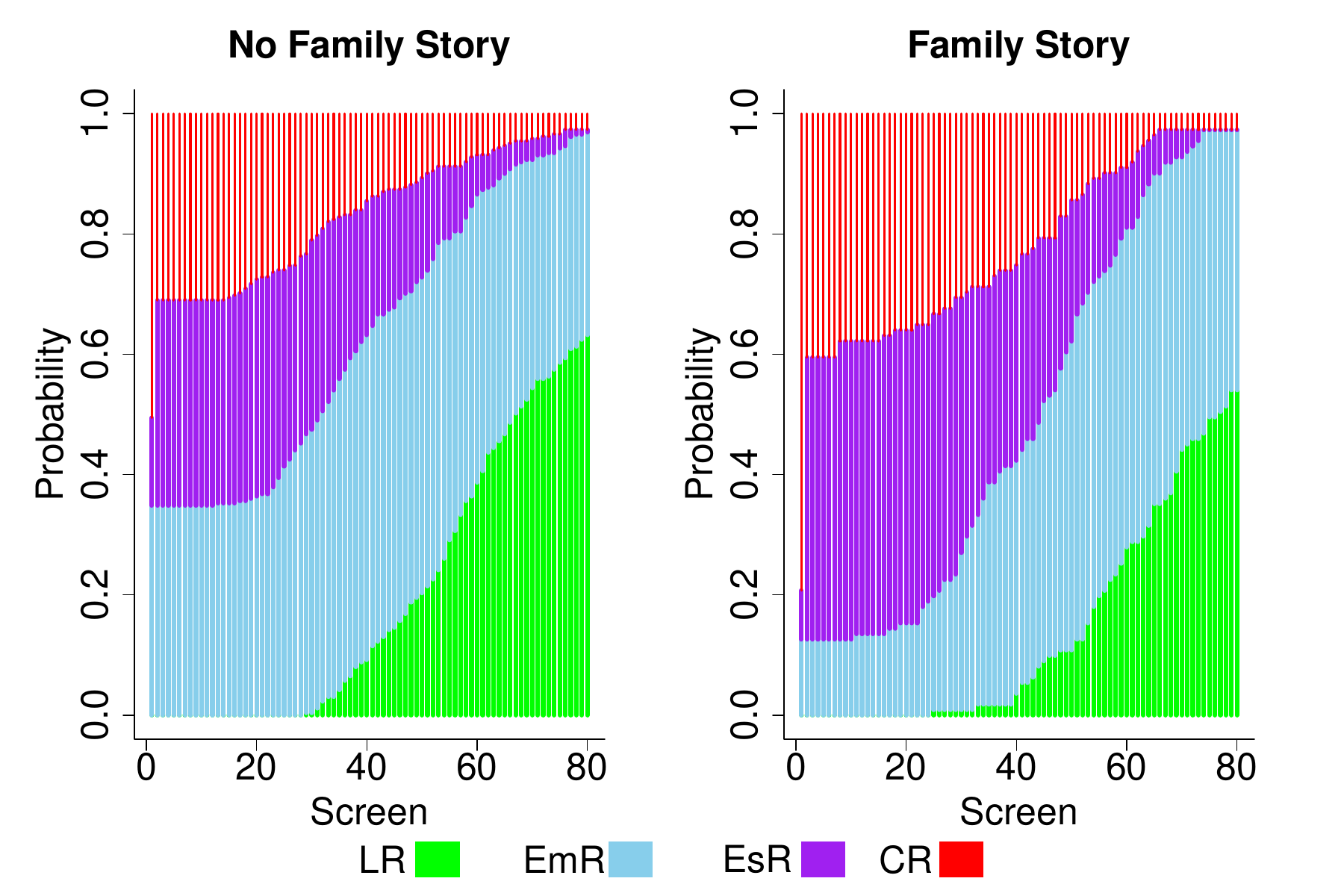}
    \caption*{\footnotesize{(a) Family Story: Overweight}}
    \label{fig:fs_ow}
\end{subfigure}
\begin{subfigure}[b]{0.48\textwidth}
    \centering
    \includegraphics[scale = 0.18]{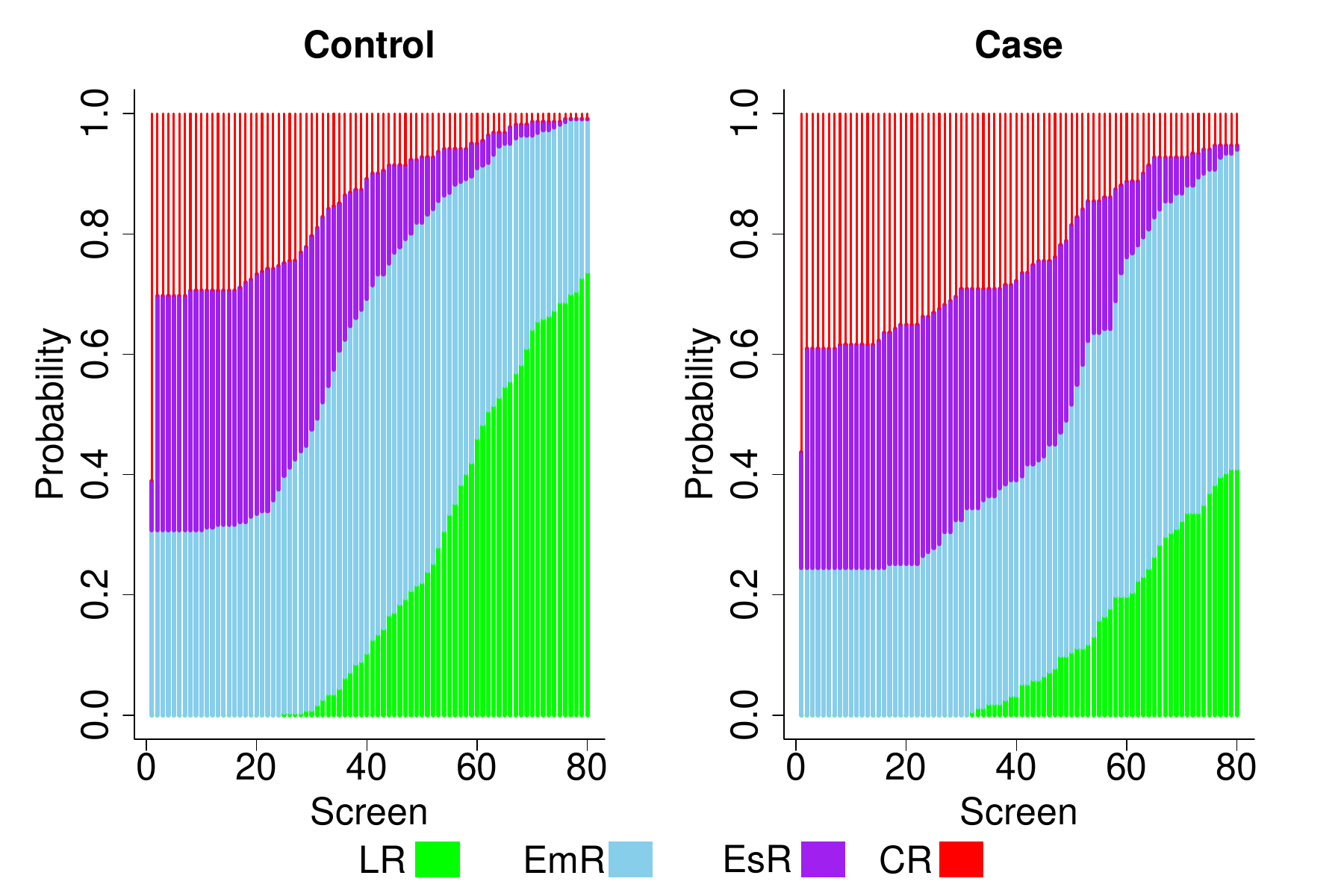}
    \caption*{\footnotesize{(b) Diagnosis Status: Overweight}}
    \label{fig:cs_ow}
\end{subfigure}

\vspace{0.8em}

\begin{subfigure}[b]{0.48\textwidth}
    \centering
    \includegraphics[scale = 0.18]{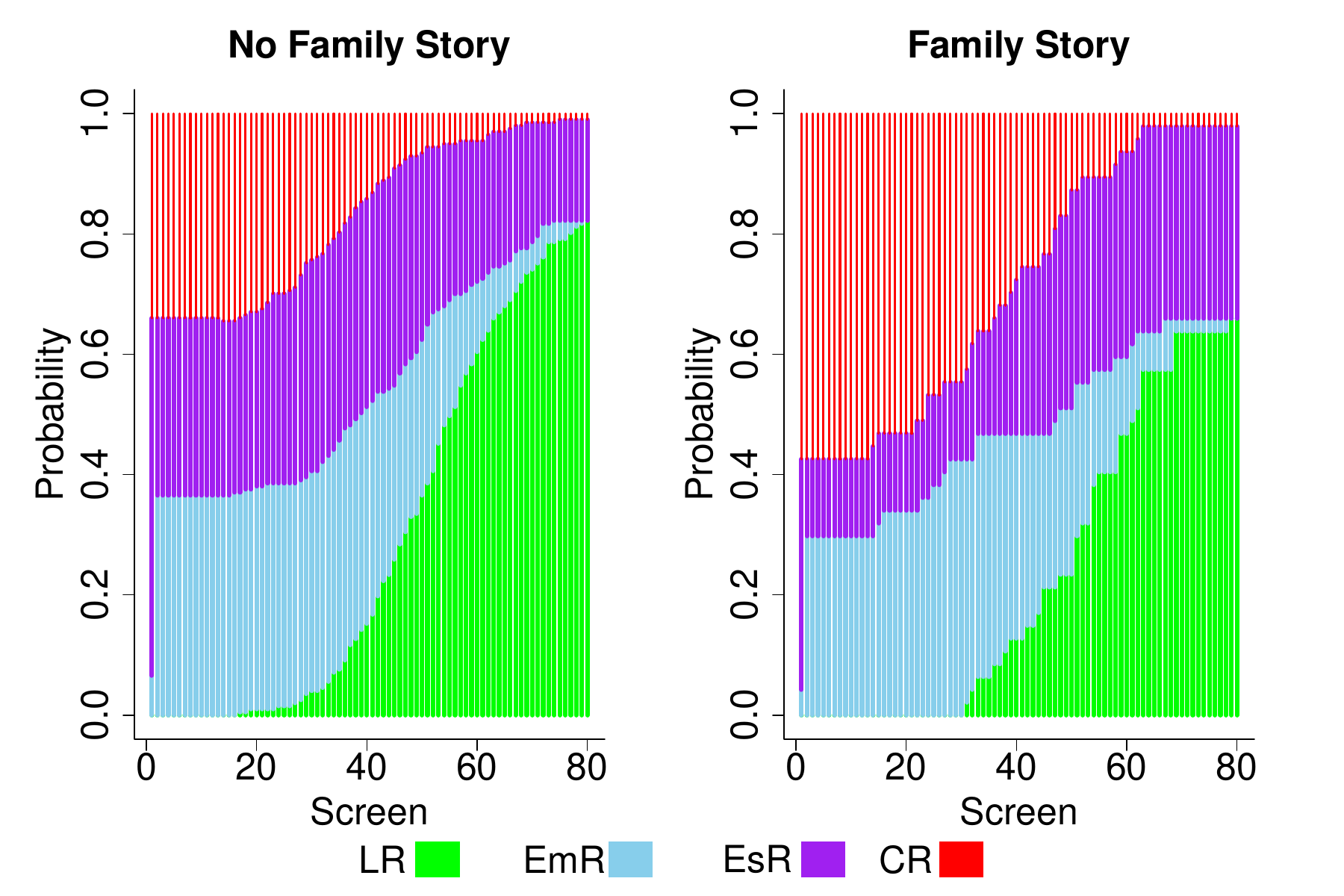}
    \caption*{\footnotesize{(c) Family Story: Obese}}
    \label{fig:fs_ob}
\end{subfigure}
\begin{subfigure}[b]{0.48\textwidth}
    \centering
    \includegraphics[scale = 0.18]{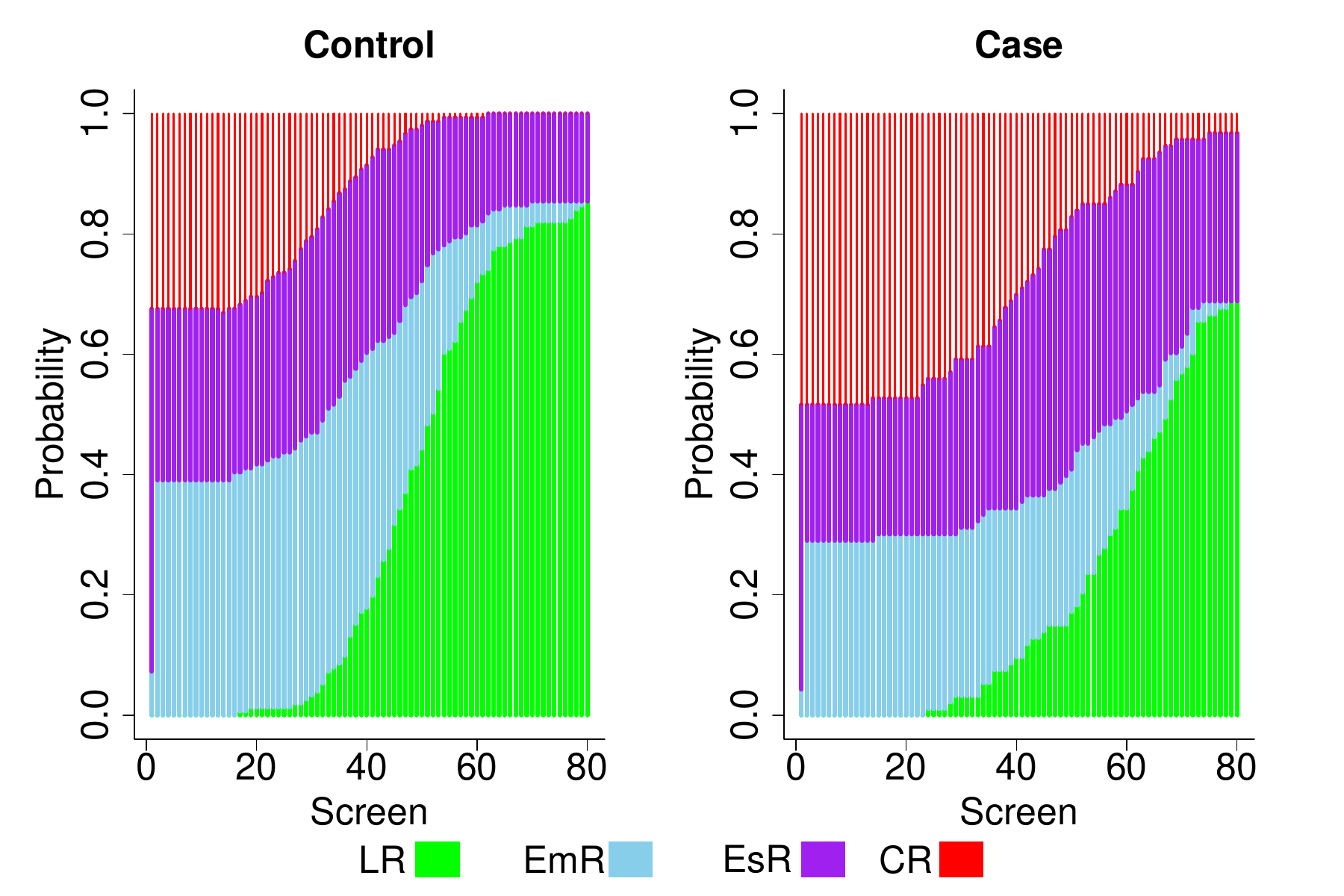}
    \caption*{\footnotesize{(d) Diagnosis Status: Obese}}
    \label{fig:cs_ob}
\end{subfigure}

\caption{Estimated latent-state proportions over screening visits for family history (left) and diagnosis status (case vs.\ control; right), shown separately for the overweight and obese BMI groups.}
\label{fig:fs_casecontrol_profiles}
\end{figure}

\begin{figure}[h!]
\centering
\begin{subfigure}[b]{0.48\textwidth}
    \centering
    \includegraphics[scale = 0.22]{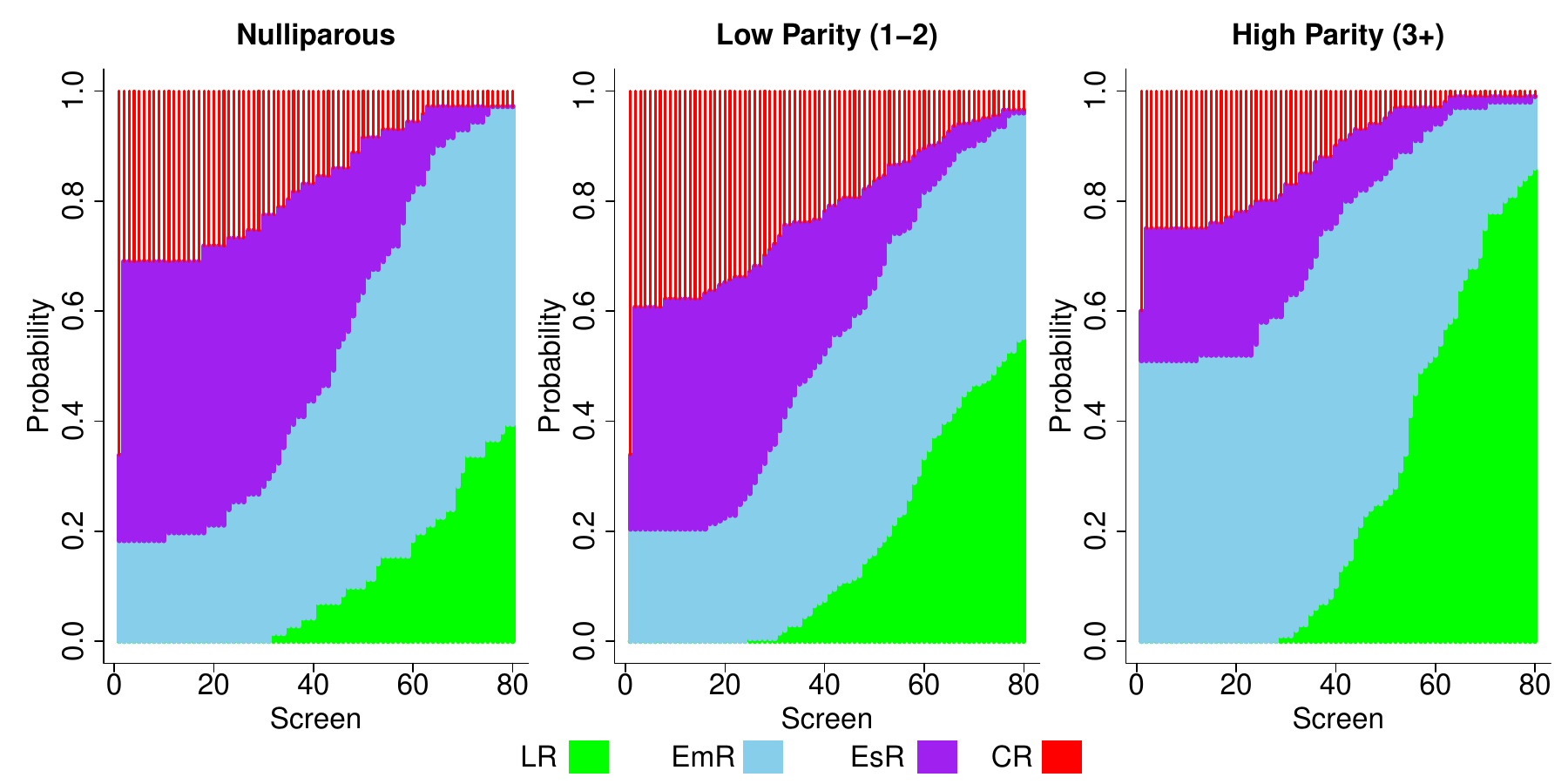}
    \caption*{\footnotesize{(a) Overweight}}
    \label{fig:sub1}
\end{subfigure}

\begin{subfigure}[b]{0.48\textwidth}
    \centering
    \includegraphics[scale=0.22]{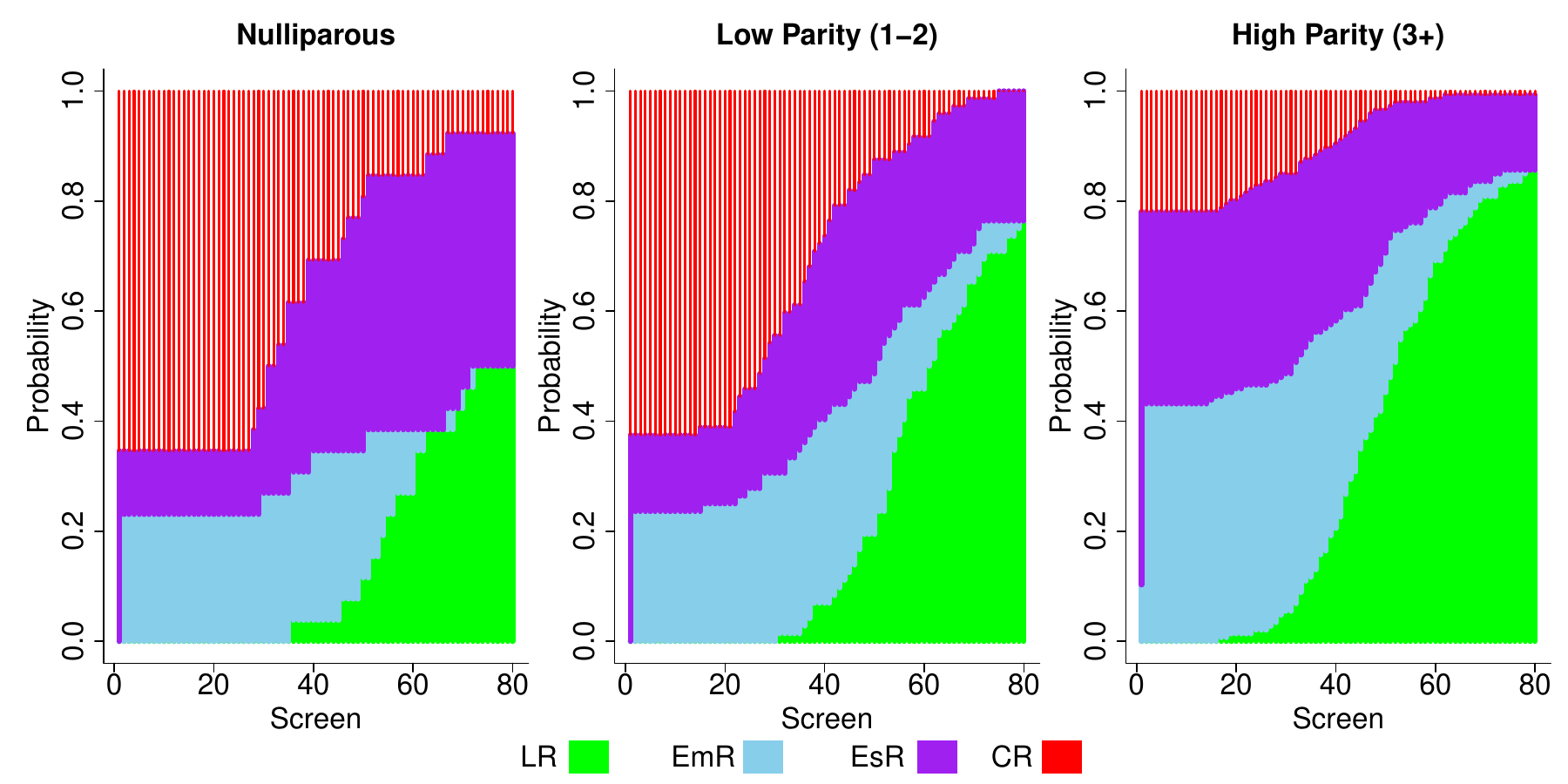}
    \caption*{\footnotesize{(b) Obese}}
    \label{fig:sub3}
\end{subfigure}

\caption{Estimated latent-state proportions over screening visits by parity groups (nulliparous, low parity [1--2], and high parity [3+]), shown separately for the overweight and obese BMI groups.}
\label{parity:transition}
\end{figure}

Building on the estimated transition parameters described in the preceding sections, we examine how the distribution of latent states evolves over follow-up. We group patients by age (33--49, 50--59, 60+), menopausal status (pre/post), family history (yes/no), parity (nulliparous, 1--2, 3+), and cancer status (control/case). Figures~~\ref{fig:age_meno_profiles}--\ref{parity:transition} display, for each subgroup and BMI category, the visit-specific proportions of patients in each latent state. In these plots, each state is represented by a distinct color; at a given visit, a larger area for a given color indicates a higher prevalence of that state in the corresponding subgroup. The aim is to illustrate whether the patterns of state occupancy over time differs between overweight and obese groups and across defined subgroups (e.g., 33-49 versus 60+).

In the overweight group, the 33--49 and 50--59 age groups show similar density transitions, yet the latent states reveal a difference that density alone would not be sufficient to show: within the same density category, the older group occupies a riskier state more often. This distinction matters over time because the 50--59 group appears healthier by the final visit. Hence, their status at the end should not be interpreted independently of the accumulated history. This same logic extends across subgroups. Across both BMI groups, pre-menopausal patients, those with a family history, those of lower parity,
and cases all show riskier transition profiles than their counterparts, spending more time in the higher-risk states---even though the difference appears small by the final visit. This is what motivates our cumulative-risk approach, which uses time spent in each latent state as the risk predictor.

\subsection{Phase~2: Detection and risk}\label{sec:res_detrisk}

Results from Phase~1 indicate that in both BMI groups, patients move from more risky latent states toward less risky latent states over follow-up (Section~\ref{sec:res_lmm_trans}). Importantly, and to remind the reader, this same direction also corresponds to movement toward less dense, and therefore more detectable, screening conditions. Hence, we capture risk as an accumulated quantity through time spent in each latent state, and separate it from detectability through the final density category. However, the detection probabilities \((p_B, p_C, p_D)\) are not identifiable from the observed diagnoses (Section~\ref{sec:det_risk}); we therefore treat them as sensitivity parameters and report three representative scenarios, ranging from a no-masking assumption to an empirically supported masking regime:
\[
\text{(i) } (p_B, p_C, p_D) = (1.0,\,1.0,\,1.0), \qquad
\text{(ii) } (0.8,\,0.6,\,0.4), \qquad
\text{(iii) } (0.8,\,0.3,\,0.2).
\]
Scenario~(i) assumes perfect detectability in every density category and serves as the baseline in
which masking is ignored, resembling the current literature; scenarios~(ii) and~(iii) introduce increasingly realistic masking of the denser categories, supported by empirical evidence~\cite{4.5} and consistent with the ordering \(p_D < p_C < p_B\). For each scenario we fit the detection--risk model of Section~\ref{sec:det_risk} and report the odds ratio and 95\% confidence interval in Table~\ref{tab:selected_state_covariate_wide_m100_b100}.

The reported estimates propagate uncertainty in the completed density histories, inferred latent states, and fitted latent Markov model, as described in Section~\ref{sec:beta_inference}. For each scenario and BMI group, we generate \(R=3\) random density completions and draw \(M=25\) posterior latent-state paths per completion within each of \(B=200\) subject-level bootstrap samples. The resulting estimates are averaged within each bootstrap sample, and percentile confidence intervals are obtained from the \(B\) replicate-level estimates. Across the two BMI groups and three detection scenarios, this requires \(2\times3\times B\times R\times M=90{,}000\) Phase~2 fits. Assumption~\ref{ass:risk} is evaluated separately using the null-model efficient score test described in Section~\ref{sec:beta_inference}, with results reported in Appendix~\ref{tab:appendix2}.

\begin{table}[h!]
\centering
\scriptsize
\setlength{\tabcolsep}{3pt}
\resizebox{\textwidth}{!}{%
\begin{tabular}{l|ccc|ccccccc}
\toprule
Group & $p_B$ & $p_C$ & $p_D$ & EmR & EsR & CR & Age & Post-menopause & Parity & Family history
 \\
\midrule
\multirow{3}{*}{Overweight} & 1 & 1 & 1 & \textbf{1.036 [1.014, 1.058]} & \textbf{1.051 [1.018, 1.079]} & \textbf{1.056 [1.032, 1.075]} & 0.976 [0.921, 1.051] & \textbf{5.956 [2.517, 13.701]} & 0.954 [0.761, 1.228] & 0.692 [0.390, 1.168]
 \\
 & 0.8 & 0.6 & 0.4 & \textbf{1.056 [1.025, 1.112]} & \textbf{1.076 [1.031, 1.123]} & \textbf{1.085 [1.050, 1.130]} & 0.972 [0.883, 1.084] & \textbf{9.344 [2.944, 47.802]} & 0.949 [0.666, 1.445] & \textbf{0.449 [0.197, 0.948]}
 \\
 & 0.8 & 0.3 & 0.2 & \textbf{1.066 [1.041, 1.102]} & \textbf{1.085 [1.041, 1.117]} & \textbf{1.090 [1.065, 1.120]} & 0.949 [0.872, 1.032] & \textbf{7.117 [2.485, 25.228]} & 0.879 [0.662, 1.175] & \textbf{0.405 [0.209, 0.750]}
 \\
\midrule
\multirow{3}{*}{Obese} & 1 & 1 & 1 & \textbf{1.056 [1.026, 1.083]} & \textbf{1.044 [1.020, 1.076]} & \textbf{1.071 [1.050, 1.103]} & 1.032 [0.944, 1.110] & 3.110 [0.996, 9.944] & 1.136 [0.908, 1.518] & 1.837 [0.819, 4.348]
 \\
 & 0.8 & 0.6 & 0.4 & \textbf{1.075 [1.037, 1.123]} & \textbf{1.065 [1.032, 1.127]} & \textbf{1.095 [1.067, 1.157]} & 1.032 [0.920, 1.134] & 3.314 [0.730, 14.817] & 1.143 [0.838, 1.756] & 1.755 [0.573, 4.605]
 \\
 & 0.8 & 0.3 & 0.2 & \textbf{1.076 [1.042, 1.115]} & \textbf{1.074 [1.044, 1.122]} & \textbf{1.098 [1.073, 1.143]} & 1.018 [0.918, 1.110] & 2.417 [0.587, 9.305] & 1.132 [0.912, 1.594] & 1.322 [0.508, 3.065]
 \\
\bottomrule
\end{tabular}%
}
\caption{Phase~2 odds ratios for the latent-state effects and adjusting covariates, estimated across \(R=3\) random density completions with \(M=25\) posterior latent-state paths per completion and \(B=200\) subject-level bootstrap replicates. Each row corresponds to one detectability scenario. Entries are odds ratios with bootstrap percentile 95\% confidence intervals; \textbf{bold} indicates an interval excluding~1.}
\label{tab:selected_state_covariate_wide_m100_b100}
\end{table}

Table~\ref{tab:selected_state_covariate_wide_m100_b100} can be read in two directions. Reading across each row, the latent-state odds ratios increase monotonically from the emergent to the
critical risk state, confirming that the states are ordered as their labels suggest: more time in a riskier state is associated with higher odds of a cancer diagnosis. Reading down the table, within each state, shows how the odds ratios change as we move from the no-masking baseline---which corresponds to the current risk analysis made in the current literature---to the empirically supported masking scenario. In every case the odds ratios increase once masking is accounted for.
   
These per-month differences may appear marginal, but they should be read on the scale at which exposure actually accumulates: patients spend on the order of a year in a given state on average, so the relevant quantity is the odds ratio compounded over twelve months. On this scale, the no-masking estimates imply that sustained occupancy of the higher-risk states raises the odds of diagnosis by about \(1.5\)--\(2.3\) times, whereas under the empirically supported masking scenario the same exposure corresponds to roughly \(2.2\)--\(3.1\) times. Because post-menopausal status is itself a significant risk factor in the overweight group, the underestimation there is largest for
post-menopausal patients, at roughly \(68\)--\(75\%\), compared with about \(41\)--\(47\%\) for pre-menopausal patients; in the obese group, the underestimation is on the order of \(25\)--\(41\%\) regardless of menopausal status.

Equally important, the effects of latent states are significant under all scenarios. Significance under no masking is what we would expect from the existing literature; had the effect been absent there, our framework would have contradicted established findings. We do not claim that masking renders an otherwise null effect significant. Rather, the contribution is to quantify how much the inferred risk grows once detectability is incorporated.

\section{Discussion}\label{sec-discussion}

This paper studies mammographic density trajectories recorded at irregular screening times and uses them to infer an underlying latent disease process. We treat this latent process as the primary component governing cancer progression, and we incorporate mammographic masking through sensitivity
parameters reflecting that a cancer can be diagnosed only once it has both developed and become detectable. This formulation decouples the dual role of mammographic density---as a risk marker and as a masking factor---by attributing risk to the latent disease structure and letting the observed density category operate primarily through detectability once that structure is accounted for.

The empirical findings support our central assumption that the latent states are informative risk indicators for breast cancer, and that the observed density carries no significant risk information once the latent structure is included. In both BMI groups, more time spent in riskier states is associated with higher odds of cancer diagnosis relative to the low-risk reference, with effect sizes that remain clinically meaningful over the one-year screening histories.

The central contribution, however, is to quantify how much these risks change once masking is taken into account. Moving from the no-masking baseline---the assumption implicit in the current
literature---to an empirically supported detectability scenario, the increase in the estimated risks is not marginal. This indicates that analyses ignoring detectability
substantially underestimate breast cancer risk. This matters most for post-menopausal overweight patients, where the underestimation rises to roughly \(75\%\), and for obese patients, where it reaches about \(40\%\). For these groups, the magnitude of the underestimation is large enough to change the clinical calculus around early intervention under the suspicion of masking, and multimodal surveillance---such as incorporating additional imaging modalities into the screening program---may therefore be appropriate and justified. A comparable degree of underestimation may not arise in normal-BMI patients; it is concentrated in these groups, where the interaction between body
mass and menopausal status plausibly drives the highest probability of a missed cancer.

Beyond the quantification of underestimation, our analysis also addresses how BMI and density category are intertwined, which complicates interpretation within a single density scale. In our analysis, we find that the resolution of observed density into latent risk states differs across BMI groups. One interpretation is that different BMI groups carry different risk profiles for the same observed densities, and that as BMI increases, this resolution occurs in lower density categories. These findings motivate BMI-stratified interpretation of density categories and further investigation into their connection.

The inferred progression patterns across patient characteristics align with established epidemiologic findings. Higher parity is associated with more favorable progression profiles, patients with a family history tend to spend more time in higher-risk states, and differences are generally more pronounced in the obese group. We also observe that younger and pre-menopausal patients in this cohort spend a larger share of follow-up in higher-risk states. However, since menopause is associated with changes in breast composition, shifts toward less dense categories should not be interpreted mechanically as improved health. Instead, our framework emphasizes accumulated history and latent-state progression. Consistently, our accumulated risk-adjusted analysis indicates that post-menopause is a significant predictor of cancer diagnosis in the overweight group. 

We acknowledge limitations that currently constrain clinical translation, including the need for larger cohorts, richer covariate information, and improved measurement completeness; these would enable more flexible covariate effects, more refined progression structures, and stronger clinical calibration. The framework also relies on modeling assumptions, including the regularization of irregular screening histories. Despite these
constraints, the proposed approach characterizes breast-density progression in relation to cancer risk and provides a concrete way to address mammographic masking. More broadly, we encourage breast cancer screening research to better leverage the longitudinal structure of density trajectories and to treat detectability as a central inferential component, moving beyond purely predictive models toward frameworks that clarify how risk accumulates, how masking affects diagnosis, and how these mechanisms vary across patient subgroups.

\bibliography{references}       

\newpage

\section*{Appendix}

\begin{table}[h]
  \centering
  \scriptsize
  \setlength{\tabcolsep}{5pt}
  \begin{tabular}{l|rr|rr}
  \toprule
  Group
  & Score statistic
  & df
  & Effective $G$
  & Bootstrap $p$-value \\
  \midrule
  Overweight & 4.910 & 2 & 1000 & 0.111 \\
  Obese     & 2.426 & 2 & 996  & 0.302 \\
  \midrule
  Combined omnibus & 7.335 & 4 & 996 & 0.142 \\
  \bottomrule
  \end{tabular}
  \caption{Empirical justification for Assumption~\ref{ass:risk}. The score is evaluated
under the null model after marginalizing the risk probability over $R=3$ random density completions and $M=25$ posterior
latent paths per completion. Bootstrap p-values are based on $G= 1000$ conditional parametric null simulations; the effective
number of converged, full-rank simulations is reported.}
  \label{tab:appendix2}
  \end{table}

\begin{table}[htbp]
\centering
\begin{tabular}{llrrrr}
\toprule
Group & $K$ & Log-likelihood & Parameters & AIC & BIC \\
\midrule
Overweight & 1 & -30971.348 & 2 & 61946.696 & 61954.534 \\
 & 2 & -12190.932 & 19 & 24419.865 & 24494.324 \\
 & 3 & -2465.912 & 46 & 5023.824 & 5204.093 \\
 & 4 & -2382.762 & 83 & 4931.523 & 5256.792 \\
 & 5 & -2379.521 & 130 & 5019.041 & 5528.498 \\
 & 6 & -2339.147 & 187 & 5052.293 & 5785.126 \\
 & 7 & -2282.633 & 254 & 5073.266 & 6068.665 \\
\midrule
Obese & 1 & -19748.845 & 2 & 39501.689 & 39508.684 \\
 & 2 & -9086.915 & 19 & 18211.831 & 18278.277 \\
 & 3 & -1512.126 & 46 & 3116.253 & 3277.123 \\
 & 4 & -1451.279 & 83 & 3068.559 & 3358.823 \\
 & 5 & -1475.492 & 130 & 3210.985 & 3665.617 \\
 & 6 & -1400.058 & 187 & 3174.116 & 3828.086 \\
 & 7 & -1447.895 & 254 & 3403.790 & 4292.070 \\
\bottomrule
\end{tabular}
\caption{Model-fit statistics for latent Markov models with $K=1,\ldots,7$ for the overweight and obese groups.}
\label{tab:appendix3}
\end{table}

\begin{table}[h!]
\centering
\scriptsize
\setlength{\tabcolsep}{3pt}
\resizebox{\textwidth}{!}{%
\begin{tabular}{l|ccc|cccccc}
\toprule
Group & $p_B$ & $p_C$ & $p_D$ & State 2 & State 3 & Age & Post-menopause & Parity & Family history
 \\
\midrule
\multirow{3}{*}{Overweight} & 1 & 1 & 1 & \textbf{1.036 [1.014, 1.058]} & \textbf{1.054 [1.033, 1.073]} & 0.981 [0.919, 1.043] & \textbf{5.677 [2.600, 12.865]} & 0.954 [0.767, 1.240] & 0.696 [0.388, 1.158]
 \\
 & 0.8 & 0.6 & 0.4 & \textbf{1.055 [1.025, 1.111]} & \textbf{1.080 [1.051, 1.130]} & 0.978 [0.882, 1.084] & \textbf{8.707 [2.955, 41.902]} & 0.943 [0.667, 1.438] & \textbf{0.455 [0.201, 0.920]}
 \\
 & 0.8 & 0.3 & 0.2 & \textbf{1.066 [1.041, 1.102]} & \textbf{1.088 [1.065, 1.116]} & 0.953 [0.876, 1.037] & \textbf{6.833 [2.423, 23.783]} & 0.877 [0.664, 1.178] & \textbf{0.407 [0.210, 0.745]}
 \\
\midrule
\multirow{3}{*}{Obese} & 1 & 1 & 1 & \textbf{1.047 [1.022, 1.075]} & \textbf{1.068 [1.049, 1.100]} & 1.015 [0.943, 1.097] & \textbf{3.494 [1.068, 11.565]} & 1.150 [0.926, 1.517] & 1.843 [0.822, 4.462]
 \\
 & 0.8 & 0.6 & 0.4 & \textbf{1.068 [1.034, 1.116]} & \textbf{1.093 [1.067, 1.156]} & 1.019 [0.918, 1.128] & 3.564 [0.696, 16.072] & 1.149 [0.881, 1.718] & 1.740 [0.592, 4.596]
 \\
 & 0.8 & 0.3 & 0.2 & \textbf{1.074 [1.043, 1.116]} & \textbf{1.097 [1.073, 1.143]} & 1.014 [0.926, 1.115] & 2.465 [0.535, 8.986] & 1.135 [0.914, 1.593] & 1.324 [0.513, 2.933]
 \\
\bottomrule
\end{tabular}%
}
\caption{Phase~2 estimates for $\textbf{K=3}$, without final density in the risk component. Point estimates average over $R=3$ random completed cohorts and $M=25$ posterior paths per completion. Intervals are empirical percentile intervals from $B=200$ subject-level bootstrap replicates; \textbf{bold} indicates an interval excluding~1.}
\label{tab:appendix4}
\end{table}

\begin{table}[h!]
\centering
\scriptsize
\setlength{\tabcolsep}{3pt}
\resizebox{\textwidth}{!}{%
\begin{tabular}{l|ccc|cccccccc}
\toprule
Group & $p_B$ & $p_C$ & $p_D$ & State 2 & State 3 & State 4 & State 5 & Age & Post-menopause & Parity & Family history
 \\
\midrule
\multirow{3}{*}{Overweight} & 1 & 1 & 1 & 0.721 [0.375, 1.145] & \textbf{1.033 [1.010, 1.059]} & \textbf{1.046 [1.023, 1.074]} & \textbf{1.052 [1.022, 1.074]} & 0.975 [0.921, 1.047] & \textbf{7.480 [2.544, 13.720]} & 0.955 [0.729, 1.243] & 0.667 [0.385, 1.214]
 \\
 & 0.8 & 0.6 & 0.4 & 0.556 [0.114, 1.067] & \textbf{1.052 [1.017, 1.095]} & \textbf{1.070 [1.038, 1.123]} & \textbf{1.079 [1.032, 1.125]} & 0.974 [0.877, 1.096] & \textbf{14.279 [3.213, 56.079]} & 0.945 [0.685, 1.421] & \textbf{0.410 [0.183, 0.907]}
 \\
 & 0.8 & 0.3 & 0.2 & 0.480 [0.077, 1.037] & \textbf{1.061 [1.031, 1.095]} & \textbf{1.077 [1.044, 1.114]} & \textbf{1.084 [1.038, 1.113]} & 0.952 [0.872, 1.047] & \textbf{11.945 [2.658, 27.920]} & 0.882 [0.665, 1.202] & \textbf{0.362 [0.188, 0.716]}
 \\
\midrule
\multirow{3}{*}{Obese} & 1 & 1 & 1 & 0.993 [0.974, 1.028] & \textbf{1.046 [1.023, 1.078]} & 1.053 [0.722, 1.107] & \textbf{1.078 [1.051, 1.116]} & 1.005 [0.928, 1.099] & \textbf{3.528 [1.121, 13.140]} & 1.183 [0.926, 1.572] & 1.733 [0.747, 4.494]
 \\
 & 0.8 & 0.6 & 0.4 & 0.988 [0.969, 1.019] & \textbf{1.068 [1.034, 1.119]} & 1.075 [0.446, 1.176] & \textbf{1.110 [1.065, 1.177]} & 1.011 [0.909, 1.132] & 3.747 [0.794, 19.587] & 1.181 [0.878, 1.759] & 1.738 [0.584, 4.893]
 \\
 & 0.8 & 0.3 & 0.2 & 0.988 [0.967, 1.012] & \textbf{1.074 [1.042, 1.118]} & 1.084 [0.706, 1.149] & \textbf{1.108 [1.074, 1.155]} & 1.008 [0.922, 1.122] & 2.549 [0.577, 9.734] & 1.156 [0.898, 1.602] & 1.325 [0.515, 3.315]
 \\
\bottomrule
\end{tabular}%
}
\caption{Phase~2 estimates for $\textbf{K=5}$, without final density in the risk component. Point estimates average over $R=3$ random completed cohorts and $M=25$ posterior paths per completion. Intervals are empirical percentile intervals from $B=200$ subject-level bootstrap replicates; \textbf{bold} indicates an interval excluding~1.}
\label{tab:appendix5}
\end{table}

\end{document}